\documentclass[
 reprint,
 amsmath,
 amssymb,
 aps,
 superscriptaddress,
 preprintnumbers,
 nofootinbib
]{revtex4-2}

\usepackage{vmargin,epsfig}
\usepackage{xcolor}
\usepackage[english]{babel}
\usepackage[toc,page]{appendix}
\usepackage{parskip}
\usepackage[hidelinks]{hyperref}
\usepackage{braket}

\usepackage{amsmath,amssymb,amsthm,amscd,color,comment}

\newcommand{\eqsp}{\, = \,}

\newcommand{\dd}{\mathrm{d}}
\newcommand{\dlog}{\dd \log}
\newcommand{\defas}{:=}

\newcommand{\Res}{{\rm Res}}
\newcommand{\Cmat}{{\bf C}}

\newcommand{\rmH}{\mathrm{H}}
\newcommand{\Dim}{\mathrm{dim}}

\usepackage{etoolbox}
\makeatletter
    \newcommand\newsubsupcommand[4]{\newcommand#1{#2\sc@subp{#3}{#4}}}
    \def\sc@subp#1#2{%
        \let\sc@subflag\undefinded%
        \let\sc@supflag\undefinded%
        \def\sc@thesub{#1}%
        \def\sc@thesup{#2}%
        \sc@proc%
    }%
    \def\sc@proc{%
        \@ifnextchar{_}{\def\sc@subflag{}\sc@mergesubs}{%
            \@ifnextchar{^}{\def\sc@supflag{}\sc@mergesups}{
                \ifdef{\sc@subflag}{}{_{\sc@thesub}}%
                \ifdef{\sc@supflag}{}{^{\sc@thesup}}%
            }%
        }%
    }%
    \def\sc@mergesubs#1#2{_{\sc@thesub#2}\sc@proc}%
    \def\sc@mergesups#1#2{^{\sc@thesup#2}\sc@proc}%
\makeatother

\newsubsupcommand{\phiL}{\varphi}{L}{}
\newsubsupcommand{\phiR}{\varphi}{R}{}
\newsubsupcommand{\psiL}{\psi}{L}{}
\newsubsupcommand{\psiR}{\psi}{R}{}
\newcommand{\abs}[1]{|#1|}

\newcommand{\namedref}[2]{\hyperref[#2]{#1~\ref*{#2}}}

\makeatletter
\def\mr@ignsp#1 {\ifx\:#1\@empty\else #1\expandafter\mr@ignsp\fi}%
\newcommand{\multiref}[1]{\begingroup%\let\protect\string%
\xdef\mr@no@sparg{\expandafter\mr@ignsp#1 \: }%
\def\mr@comma{}%
\@for\mr@refs:=\mr@no@sparg\do{\mr@comma\def\mr@comma{,\,}\ref{\mr@refs}}%
\endgroup}
\makeatother
\renewcommand{\eqref}[1]{(\multiref{#1})}

\newcommand{\be}{\begin{equation}}
\newcommand{\ee}{\end{equation}}
\newcommand{\bea}{\begin{eqnarray}}
\newcommand{\eea}{\end{eqnarray}}
\newcommand{\bei}{\begin{itemize}}
\newcommand{\eei}{\end{itemize}}

\newcommand{\unipd}{Dipartimento di Fisica e Astronomia, Universit\`a degli Studi di Padova,
Via Marzolo 8, I-35131 Padova, Italy.}

\newcommand{\pdinfn}{INFN, Sezione di Padova,
Via Marzolo 8, I-35131 Padova, Italy.}

\definecolor{green1}{HTML}{244819}
\definecolor{cyan1}{HTML}{37cdaa}
\definecolor{blue1}{HTML}{5d7ac4}
\definecolor{red1}{HTML}{d0482a}
\definecolor{purple1}{HTML}{845ea8}
\definecolor{orange1}{HTML}{e07229}
\begin{document}

%\hfill \small{HU-EP-24/33-RTG}
\preprint{HU-EP-24/33-RTG}

\title{Gluing via Intersection Theory}

\newcommand{\higgs}{Higgs Centre for Theoretical Physics, School of Physics and Astronomy The University of Edinburgh, Edinburgh EH9 3FD, Scotland, UK}

\author{Giulio~Crisanti}\email{g.crisanti@ed.ac.uk}
\affiliation{
\unipd,
\pdinfn
} 
\affiliation{
\higgs
} 

\author{Burkhard~Eden}\email{eden@math.hu-berlin.de}
\affiliation{%
Institute f\"ur Mathematik und Physik, Humboldt-Universit\"at zu Berlin, Zum gro{\ss}en Windkanal 2, 12489 Berlin, Germany.}

\author{Maximilian~Gottwald}\email{gottwalm@physik.hu-berlin.de}
\affiliation{%
Institute f\"ur Mathematik und Physik, Humboldt-Universit\"at zu Berlin, Zum gro{\ss}en Windkanal 2, 12489 Berlin, Germany.}

\author{Pierpaolo~Mastrolia}\email{pierpaolo.mastrolia@unipd.it}
\affiliation{
\unipd,
\pdinfn}

\author{Tobias~Scherdin}\email{scherdit@physik.hu-berlin.de}
\affiliation{%
Institute f\"ur Mathematik und Physik, Humboldt-Universit\"at zu Berlin, Zum gro{\ss}en Windkanal 2, 12489 Berlin, Germany.}

\date{\today}

\begin{abstract}
Higher-point functions in $\mathcal{N}=4$ super Yang-Mills theory can be constructed using integrability by triangulating the surfaces on which Feynman graphs would be drawn. It remains hard to analytically compute the necessary re-gluing of the tiles by virtual particles.

 We propose a new approach to study a series of residues encountered in the two-particle gluing of the planar one-loop five-point function of stress tensor multiplets. After exposing the twisted period nature of the integral functions, we employ intersection theory to derive canonical differential equations and present a solution.

\end{abstract}

\maketitle
\section{Introduction}
\begin{figure}[b]
\hskip 0.5 cm \includegraphics[width = \linewidth]{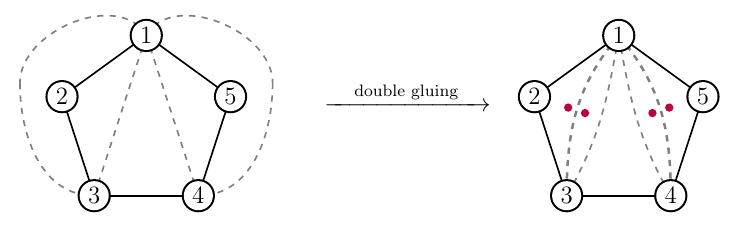} 
\caption{The left part shows one possible triangulation of the correlation function of five stress tensor multiplets by six hexagons. In the right panel a double gluing of the tessellated graph is depicted in which the front side is dressed by virtual particles.}
\label{fig:glue}
\end{figure}
\noindent
The maximally supersymmetric non-Abelian gauge theory in four dimensions, ${\cal N} \eqsp 4$ super Yang-Mills (SYM) theory, is conformally invariant even at the quantum level \cite{PhysRevLett.45.1063,CASWELL1981152,mandelstam1982light,BRINK1983401,BRINK1983323,HOWE1984125}. Therefore, natural \emph{observables} are correlation functions of gauge invariant composite operators. The latter are classified by their spin, flavour, and scaling dimension.

The planar Feynman graphs causing quantum corrections to the dimension label can be viewed as a spin chain Hamiltonian \cite{Minahan:2002ve,Beisert:2004hm,Beisert:2005fw,Janik:2006dc}. Bethe ansatz techniques have led to spectacular progress on the planar spectrum problem: e.g. systematic weak- and strong-coupling expansions can be derived for the planar \emph{cusp anomalous dimensions} and numerical interpolation between these two extremes is possible \cite{BES1,BES2,Benna:2006nd}. 

These methods can be used to study higher-point quantities as well. The first such application concerned planar on-shell scattering amplitudes \cite{Basso:2013vsa,Basso:2014koa,Basso:2014nra}. In the weak-coupling regime, series of residues obtained from this integrable system could be matched with ansätze of special functions up to very high loop order, outmanoeuvring Feynman graph and even unitarity based computations. We emphasize that an analytic evaluation of the resulting sum-integrals has been achieved only in very few cases, cf. \cite{Papathanasiou:2013uoa,Cordova:2016woh,Lam:2016rel} where hypergeometric functions were obtained and evaluated in terms of generalised polylogarithms.

Secondly, higher-point functions of gauge-invariant composite operators can be built from hexagonal patches \cite{BKV,cushions,shotaThiago} defined by elements of the Bethe ansatz for the anomalous dimensions. In a similar manner to the cusp anomalous dimensions, \emph{integrability} works best in special limits. For instance, a certain four-point function has been determined to all orders in the Yang-Mills coupling and the rank of the gauge group \cite{Bargheer:2019exp,Bargheer:2019kxb,Kostov:2019stn}. 

In order to exactly reproduce quantum field theory, the hexagon tiles must be \emph{glued} by virtual particles. Each of these comes with a Mellin-like integral and two counters akin to a radial and a magnetic quantum number. They scatter on a hexagon by the \emph{bound state $S$ matrix} \cite{glebBound} dressed with a scalar factor \cite{BES1,BES2,glebSergey}. The lowest generic process of this type contributes to the one-loop five-point function of stress-tensor multiplets \cite{shotaThiago2,usFivePoints} sketched in Figure~\ref{fig:glue}. Away from any limit, the analytic evaluation of such processes remains an open problem.

In this letter, we employ the recently developed methods of intersection theory for twisted de Rham cohomology \cite{interTheo1,Mastrolia:2018uzb,fatIntersection2} to leverage the period integral structures underpinning the entire bound state scattering, to help overcome the computational challenges in the program of \cite{usFivePoints}. With respect to the agenda of re-summing the most general gluing contributions, this is certainly only a first step, though - as we believe - a significant one. We show, for the first time, that all the tools of intersection theory become available for gluing computations and for gluing-based methods: Through the usage of intersection numbers, integral decompositions as well as systems of Pfaffian differential equations can be derived by projection - and eventually solved.

\section{Double Gluing Process} 
\label{theProblem}
\noindent
The virtual particles in gluing processes are \emph{bound states} formed out of $K$ elementary constituents, also called the \emph{length} of the composite. The constituents can be two-component fermions, or bosons with a two-component flavour index. All constituents of such a state move along the spin chains as a block with one common rapidity. The entire composite is an antisymmetrised tensor product. This can contain an arbitrary number of fermions due to their statistics, but at maximum one boson of any given flavour. Hence there are four types of states, having no boson, one of $\phi^1, \, \phi^2$, or both.

We limit the scope of the present discussion to the contribution to the process in Figure~\ref{fig:glue} \cite{shotaThiago2,usFivePoints} in which both states are of the form
\begin{equation}
\begin{aligned} \label{bsOne}
&|K,k\rangle_1 \eqsp (\psi^1)^{K-k-1}(\psi^2)^k\phi^1, \\
&~~~~~~K,\, k\in \mathbb{N}_0,\,K>k\geq 0 \, .
\end{aligned} 
\end{equation}
Here the magnetic quantum number $k$ counts the number of up-spins, and $\phi^1$ is a single scalar constituent. 

Their scattering on the central tile takes the form
\begin{align}
\begin{split}
\mathbb{S}|K,k;L,l\rangle = \sum_{n=0}^{k+l}\mathcal{X}^{k,\,l}_n|K,n;L,k+l-n\rangle
\end{split}
 \end{align}
where $L, \, l$ are the quantum numbers of the second particle and $\mathbb{S}$ is the scattering matrix as an operator. The relevant matrix element is $\mathcal{X}$.

The details of the formalism \cite{BKV} imply diagonal scattering (so $n \eqsp k$) because the outer two hexagons do not contain any other particles. Up to a rescaling the diagonal $\mathcal{X}$-elements are
\begin{equation}\begin{aligned}\label{eq:Xmatrix}
\hat{\mathcal{X}}^{k, \,l}_k& =\sum_{m=0}^{k}
 \tbinom{k}{k - m} \tbinom{l}{k - m}\tfrac{(-1)^{-k + l + m} \Gamma[L - l] \Gamma[K - m] }{ \Gamma[K - k] \Gamma[L - k - l + m]}\\
&* \tfrac{\Gamma[ i u^- - i v^- -k + l ] \Gamma[ i v^- - i u^+ -k - l ] \Gamma[1 - i u^+ + i v^+]}{ \Gamma[ i u^- - i v^- -m] \Gamma[1 - m - i u^+ + i v^+] \Gamma[ iv^- - i u^+ ]} \, .
\end{aligned}
\end{equation}
This matrix part of the scattering is completed by the BES phase in special kinematics \cite{BES2,glebSergey}
\begin{equation}\label{eq:BESPhase}
\Sigma^{KL}=\frac{\Gamma[1+iu^-]\Gamma[1-iv^+]\Gamma[1- iu^+ + iv^-]}{\Gamma[1 -iu^+]\Gamma[1+iv^-]\Gamma[1+iu^- - iv^+]}\,,
\end{equation}
with $u^{\pm}=u\pm\tfrac{i}{2}K$, $v^{\pm}=v\pm\tfrac{i}{2}L$. Here, $u, \, v$ are \emph{rapidities} for the movement of the bound states along the spin chain.

A full evaluation of the gluing process in Figure \ref{fig:glue} not only involves sums over the quantum numbers of the two states located on the edges $1-3$ and $1-4$ of the central hexagon tile and integrals over their respective rapidities, but also a sum over the four possible types of flavours on either occupied edge \cite{shotaThiago2,usFivePoints}. Processes with the other flavours of particles are similar but involve more complicated scattering matrices ${\cal Y, \, Z}$ composed of two or three ${\cal X}$ matrices with shifted indices respectively, and some additional rational coefficients \cite{glebBound,DeLeeuw:2020ifb}. 

The $\mathcal{X}$-part of the double gluing process in Figure \ref{fig:glue} can be expressed by the sum-integral \cite{shotaThiago2,usFivePoints}
\begin{equation}\label{eq:integral}
\begin{aligned}
I_{\mathcal{X}}=\hspace{-2pt}\sum_{K,\,L=1}^{\infty}\hspace{-3pt}\sum_{k,\,l=0}^{K-1,L-1}\hspace{-3pt}\int\tfrac{du\,dv}{4\pi}\tfrac{K\,L\,g^2(v^+-u^-)\, \Sigma^{KL}\,\hat{\mathcal{X}}^{k,l}_k\, W}{(u^-)^2 (v^+)^2u^+v^-(u^+-v^-)}
\end{aligned}
\end{equation}
with $W=z^{-iu^+-k}b^{-iu^-+k} a^{iv^++l} y^{iv^--l}$.

As a simplification we restrict the outer kinematical variables of the five-point function to be located on a plane. In doing so we reduce the number of independent cross ratios to four. They are parametrised by 
\begin{equation}\begin{aligned}\label{cross1234} 
b z \eqsp \frac{x_{12}^2 x_{34}^2}{x_{13}^2 x_{24}^2} \, , \qquad (1-b)(1-z) \eqsp \frac{x_{14}^2 x_{23}^2}{x_{13}^2 x_{24}^2}
\end{aligned}\end{equation}
with $x_{ij} \eqsp x_i - x_j$, and similarly for $1/a$, $1/y$ w.r.t. the points $\{1,3,4,5\}$.

In order to evaluate \eqref{eq:integral}, we close the integration contours for $u$ and $v$ in the upper and lower half-plane, respectively. Furthermore, the integration containing a double pole must be taken first. Since $I_{\mathcal{X}}$ includes a pole of second order in both variables $u$ and $v$, it is convenient to split \eqref{eq:integral} into two parts using the numerator $(v^+-u^-)$ to cancel one of the double poles for either variable.

The integration over $u$ and $v$ results in the following four summation processes:
\begin{equation}
I_{\mathcal{X}}= (S_1 + S_2) + S_{W} + S_{mes} + S_{mat}
\end{equation}
where $S_1$ and $S_2$ denote series of residues created by a derivative from a double pole in $v$ and $u$ respectively, falling onto the third denominator $\Gamma$ function in the dressing phase \eqref{eq:BESPhase}. Here the residues of the poles of the resulting digamma function are picked to localize the second variable. Both are sixfold sums.

In the remaining simpler parts $S_W, \, S_{mes}, \, S_{mat}$, both variables are frozen using the residues from $u^-, \, v^+$, and the derivative from the double pole acts on $W$, $\, u^+ v^-$ and $\mathcal{X} \, \Sigma^{KL} / (u^+-v^-)$, respectively. These three cases have been discussed in \cite{usFivePoints}. In this work we take a closer look at $S_1$ and $S_2$ which are explicitly given by
\begin{equation}\label{defS12}
\begin{aligned}
S_1 & =  \sum a^{\sigma_{lm}} b^{\sigma_{jkm}}  y^{\sigma_{Lm}} z^{\sigma_{jK}} \frac{\sigma_{kKm}}{j \sigma_{jkKm}}  \\
& *  \frac{\Gamma_{1km} \Gamma_{1lm} \Gamma_{Km} \Gamma_{Lm} \Gamma_{jKLm} \Gamma_{1jklLM}}{\Gamma_{1k} \Gamma_{1l} \Gamma_K \Gamma_L \Gamma_{1m}^2 \Gamma_{1lLM} \Gamma_{1jKm} \Gamma_{1jkLM}} \, ,  
\\[8pt]
S_2 & =  \sum a^{\sigma_{jlm}} b^{\sigma_{km}}  y^{\sigma_{jLm}} z^K \frac{\sigma_{lLM}}{j \sigma_{jlLM}}  \\ 
& *  \frac{\Gamma_{1km} \Gamma_{1lm} \Gamma_{Km} \Gamma_{Lm} \Gamma_{jKLm} \Gamma_{1jkKm}  \Gamma_{1jklLM}}{\Gamma_{1k} \Gamma_{1l} \Gamma_K \Gamma_L \Gamma_{1m}^2 \Gamma_{1kKm}  \Gamma_{1jKm} \Gamma_{1jkLM} \Gamma_{1jlLM}} 
\end{aligned}
\end{equation}
with $\sigma_{ij\ldots} = i+j+\ldots, \, \Gamma_{ij\ldots} = \Gamma[i+j+\ldots] $, $M=2m$. The summation ranges are $j,K,L \geq 1; k,l,m \geq 0$. 

Comparing the two series \eqref{defS12} at leading orders we find
\begin{equation}\begin{aligned}\label{theFeature}
S_2 \eqsp S_1(a \, \leftrightarrow \, b, \, y \, \leftrightarrow \, z) \, .
\end{aligned}\end{equation}
The underlying flip symmetry of the $\mathcal{X}$ matrix is non-manifest. Since future two-loop computations will contain higher poles in both $u^-$ and $v^+$, it will be vital to be able to handle the bound state scattering in all situations without appealing to the exchange symmetry. Our example illustrates the difficulties this causes.

The individual sums in \eqref{defS12} are all of hypergeometric $_{p+1}F_p$ type. If $p \eqsp 0$ this is a geometric series, else we may replace it by the Euler integral representation
\begin{equation}\begin{footnotesize}\begin{aligned}\label{Euler}
_2F_1[\{c,d\},\{e\},z] \rightarrow  \frac{\Gamma[d]^{-1}\Gamma[e]}{ \Gamma[e-d]} \int_0^1 ds \, s^{d-1} \frac{(1-s)^{e-d-1}}{(1 - s z)^{c}}  , 
\end{aligned}\end{footnotesize}\end{equation}
or $p$-fold Euler integrals for $_{p+1}F_p$. We can iteratively apply this substitution to eventually arrive at an integral representations of the two sums: 
\begin{equation}\begin{aligned}\label{eq:integrals}
S_1&=\int_0^1 ds \int_0^1 dt \, \frac{b y z^2 \, (1-t)}{q_1|_{t \eqsp 0} \, q_1|_{t \eqsp 1} \, q_1} \, ,\\
S_2&=I_1+I_2 \, , \\
I_1&=\int_0^1 ds \int_0^1 dt \, \frac{n_1(s,t)}{q_2(s,t)q_3(s,t)} \, , \\
I_2&=\int_0^1 dr \int_0^1 ds \int_0^1 dt \, \frac{n_2(r,s,t)}{\left(q_4(r,s,t)\right)^2}
\end{aligned}\end{equation}
The polynomials $d_j, \, p_j, \, n_j, \, q_j$ occurring here and in later equations are given in Appendix~\ref{App:polynomials}.

The integral representation of $S_1$ can be evaluated directly:
\begin{equation}\begin{aligned}
S_1 \eqsp \frac{y z}{d_1} \, s_{1,1} + \frac{y z}{d_2} s_{1,2}
\end{aligned}\end{equation} 
The symbols \cite{Goncharov:1998kja,Goncharov:2010jf} $s_{1,1}, \, s_{1,2}$ of the respective dilogarithm functions are given by
\begin{equation}\begin{footnotesize}\begin{aligned}
s_{1,1} & =  d_2 \otimes \frac{(1-a-b)(1-y)(1-z)}{(1-a)(1-b)(1-y-z)} \\&
+ d_3 \otimes \frac{(1-a)(1-y-z)}{(1-y)(1-a-b)} 
+ (b-z) \otimes \frac{(1-b)}{(1-z)} \\&
+ (1-a) \otimes \frac{(1-a-b)}{(1-a)(1-z)} 
+  (a b) \otimes \frac{(1-a)(1-b)}{(1-a-b)} \\&
- (1-y) \otimes \frac{(1-y-z)}{(1-b)(1-y)}  
- (y z) \otimes \frac{(1-y)(1-z)}{(1-y-z)} 
\end{aligned}\end{footnotesize}\end{equation}
and
\begin{equation}\begin{footnotesize}\begin{aligned}
s_{1,2}  = & \frac{(1-a-b)}{(1-a)} \otimes (1-z) + (1-z) \otimes \frac{(1-a-b)}{(1-a)}  \\
 - & \frac{(1-y-z)}{(1-y)} \otimes (1-b) - (1-b) \otimes \frac{(1-y-z)}{(1-y)} - s_{1,1} \, .
\end{aligned}\end{footnotesize}\end{equation}
However, in the case of $I_1$ and $I_2$ direct integration is more challenging. We resort to methods from intersection theory \cite{interTheo1,Mastrolia:2018uzb} instead which will allow for a systematic computation of these two integrals.

\section{Intersection Theory and Gluing Integrals} \label{introIntersect}

\noindent
{\bf Twisted Cohomology Groups.} \textit{Twisted co-homology} theory 
studies the algebraic properties of {\it twisted period integrals} of the form
\begin{align}
    I &= \int_{\mathcal{C}_R} \! u \ \phiL \defas \langle \phiL | \mathcal{C}_R ]
    \>,
    \label{eq:integraldef}
\end{align}
where $\phiL$ is a rational $n$-form, say 
$\phiL = {\widehat \varphi}_L(z) \, \dd \mathbf{z} \defas {\widehat \varphi}_L(z) \,
\dd z_1 \wedge \ldots \wedge \dd z_n$,
and $u$ is a multivalued function, which, by requirement, vanishes at the boundary of the integration domain ${\cal C}_R$, {\it i.e.} 
$u(\partial {\cal C}_R)=0$.
Owing to the latter condition, any $n-1$ form $\phi_L$ obeys the integration-by-part (IBP) identity 
\begin{equation}
    \int_{\mathcal{C}_R} \! \dd (u \, \phi_L) =  
\int_{\mathcal{C}_R} \! u \, \nabla_\omega \, \phi_L  =
\langle \nabla_\omega \, \phi_L | \mathcal{C}_R ] = 0\, ,
    \label{eq:ibpi}
\end{equation} 
where we introduce 
the covariant derivative $ \nabla_\omega \defas \dd + \omega \>, $ with
\begin{align}
     \omega \defas \dlog(u)
    = \sum_{i=1}^n \widehat{\omega}_i \, \dd z_i \>,
    \ \text{and} \quad 
    \widehat{\omega}_i \defas \partial_{z_i} \log(u) \>.
    \label{eq:omega_def}
\end{align}
Within twisted de Rham theory, any $n$-form $\phiL$
is an element of the $n$\textsuperscript{th} twisted cohomology
group $\rmH^{n}_\omega\>$, defined as the vector space of equivalence classes of closed,
modulo exact, $n$-forms
$\langle \phiL | \sim \langle \phiL + \nabla_\omega \phi_L| \,.$
Thus, upon integration, any member of each equivalence class produces the same result: 
$
\langle \phiL + \nabla_\omega \phi_L| \mathcal{C}_R ] 
=\langle \phiL |\mathcal{C}_R ] =  I\,.
$
The dimension \mbox{$\nu$} of the co-homology group 
\begin{align}\label{eq:leepom}
    \nu
    \defas \Dim\left( \rmH^{n}_{\omega} \right)
\end{align}
corresponds to the number of independent equivalence classes of $n$-forms $\{ \langle e_i | \}_{i=1, \cdots ,\nu}$, which can be considered as generators of ${\rmH_{\omega}^n}$. It also counts the number of master integrals
which is determined by the number of critical points of the Morse height function $\log (\abs{u})$ \cite{Lee:2013hzt} and can be computed by {\it counting} the number of solutions of the (zero dimensional) system
\begin{equation}
    \nu = \text{\# of solutions to}\,\,\, (\omega_i = 0)\,.
    \label{eq:leepom2}
\end{equation}
for $i=1,\ldots, n$.

{\bf Intersection Numbers.} The (twisted cohomology) intersection number $\langle \varphi_L | \varphi_R \rangle$ \cite{cho1995,matsumoto_1998} is an inner product between elements of a twisted cohomology group
$\varphi_L \in \rmH^{n}_{\omega}$,
and its respective dual
$\rmH^{n}_{-\omega}$,
defined by sending $u \to u^{-1}$, $\omega \to -\omega$.

As for any vector space, this inner product allows for the decomposition of any element onto a basis simply  by projection using the \textit{master decomposition formula}~\cite{Mastrolia:2018uzb,Frellesvig:2019kgj}
\begin{align}
    \langle \phiL |
    =
    \sum_{i=1}^{\nu} c_i \langle {e}_i |
    \>,
    \quad
    \text{with}
    \quad
    c_i = \langle {\phiL} \,|\, {h}_j \rangle (\Cmat^{-1})_{ji}
    \>,
    \label{eq:masterdecomposition}
\end{align}
where 
$\{ \langle e_i | \}_{i=1, \cdots ,\nu}$ and 
$\{ | h_i \rangle \}_{i=1, \cdots ,\nu}$ are a bases 
of $\rmH^{n}_{\omega}$ and $\rmH^{n}_{-\omega}$, respectively. The elements of the \textit{$\Cmat$-matrix}
\begin{align}
    \Cmat_{ij} \defas \langle {e}_i \,|\, {h}_j \rangle
    \label{eq:firstCdef}
\end{align}
are intersection numbers between the basis and dual basis elements. Any Euler/Mellin/Feynman integral as in \eqref{eq:integraldef} can thus be decomposed in terms of master integrals $J_{i}$ as
\begin{align}
    I = 
      \langle \phiL | \mathcal{C}_R ]\
    = \sum_{i=1}^{\nu} c_i \langle e_i | \mathcal{C}_R ]
    = \sum_{i=1}^{\nu} c_i \, J_i \, ,
\end{align}
where the master integrals 
$J_i = \langle e_i | \mathcal{C} \big]$ correspond to the integrals generated by the basis forms $\langle e_i|$.

\noindent
{\bf Differential Equations}.
The master decomposition formula in \eqref{eq:masterdecomposition}
can be used to derive the system of differential equations obeyed by the master forms $\langle e_i|$. Assume that $u$ depends on an external variable, say $x$. Then 
\begin{eqnarray}
\partial_x \langle e_i | = \langle \nabla_{\sigma_x} \, e_i | 
= \Omega_{ij} \, \langle e_j | \ ,
\label{eq:maserdeqsystem}
\end{eqnarray}
where $\sigma_x := \partial_x \log(u)$ and $\nabla_{\sigma_x} := \partial_x + \sigma_x$. The elements of the matrix ${\bf \Omega}$ can thus be computed in terms of intersection numbers
\begin{align}\label{eq:intersectDEQ} 
\Omega_{ij} = \langle \nabla_{\sigma_x} \, e_i | h_k \rangle \, ({\bf C}^{-1})_{kj}\,.
\end{align}

Since integrals are obtained by pairing forms and integration contours, the connection matrix ${\bf \Omega}$ is also the matrix for a system of differential equations obeyed by the master integrals $J_i$
\begin{eqnarray}
\partial_x \, J_i = \Omega_{ij} \, J_j \, . 
\end{eqnarray}
Correspondingly, the computation of intersection numbers allows for the construction of differential equation systems obeyed by the master integrals. 

{\bf Gluing Integrals.}
The integrals $\hat I_1$ and $\hat I_2$ we consider in this work admit a twisted period representation of the form \eqref{eq:integraldef}, where the integration domain is  
\mbox{${\cal C}_R = [0,1] \times [0,1]$}.

The twist $u$ in general takes the form
\begin{equation}\begin{aligned}
u \eqsp s^\gamma (1-s)^\gamma t^\gamma (1-t)^\gamma q_1^\gamma \ldots q_n^\gamma \, ,
\end{aligned}\end{equation}
where $\varphi_L \in {\rm H}^2_\omega$ is a 2-form
\begin{equation}\begin{aligned}
\varphi_L \eqsp \hat \varphi_L(s,t) \, ds \wedge dt \ .
\end{aligned}\end{equation}
${\hat \varphi_L}$ is rational function in the variables $s$ and $t$, whose numerator may contain polynomials in $s$ and $t$. The denominator is built from factors belonging to the twist, namely 
\begin{equation}
\{s, (1-s), t, (1-t), q_1, \ldots, q_n\}
\ . 
\end{equation}
The singularities arising at the integration boundaries are analytically regulated by the generic exponent $\gamma$ appearing in the twist. This ensures that the condition $u(\partial \, \mathcal{C}_R)=0$ is always satisfied.

Deriving systems of differential equations for $I_1$ and $I_2$ requires the evaluation of intersection numbers for generic meromorphic differential $2$-forms. These are hereby computed following the fibration-based algorithm (for $n$-forms) \cite{Mizera:2019gea,fatIntersection2,Frellesvig:2019uqt}. We describe this algorithm in detail in Appendices \ref{sec:1_form_comp} and \ref{sec:2_form_comp}, and refer the reader to \cite{Brunello:2023rpq,Brunello:2024tqf} for recent developments.

A discussion of the details regarding this specific calculation is additionally provided in Appendix \ref{sec:comp_details}.

\section{The two-parameter integral $I_1$} \label{secIntI2}
By partial fractions in $s$, the integral $I_1$ in \eqref{eq:integrals} decomposes into two individually finite integrals 
\begin{align}
I_1=\tilde I_1 + \hat I_1 \, .
\end{align}
$\tilde I_1$ can be easily integrated directly. The expression is provided in the \emph{ancillary files}. We now consider $\hat I_1$, which is given by
\begin{equation}
\begin{aligned}
\label{defHatI2}
\hat I_1 = \int_{\mathcal{C}_R} u_1(s,t) \, \varphi_1(s,\,t) = \langle \varphi_1(s,\,t) | \mathcal{C}_R]~~~\\
\text{with }~~u_1(s,t)\eqsp s^\gamma (1-s)^\gamma t^\gamma (1-t)^\gamma p_1^\gamma p_2^\gamma q_2^\gamma,\\
\varphi_1(s,\,t)=\frac{y^2 (1-Z) \, (1-t) \, n_3(s,t)}{p_1(t) \, p_2(t) \, q_2(s,t)} \, ds \wedge dt,
\end{aligned}
\end{equation}
in the $\gamma\rightarrow0$ limit.

\textbf{Bases.} By using \eqref{eq:leepom2} we find $\nu=13$, hence $\hat I_1$ can be decomposed in terms of a basis of 13 master integrals, whose corresponding basis of master 2-forms 
$\langle e_i | = {\hat e}_i 
\, \dd s \wedge \dd t $ 
is chosen to be
\begin{equation}\begin{aligned}
 \{ {\hat e}_i \}_{i=1}^{13}
 = \Bigl\{ &\frac{1}{q_2}, \frac{1-t}{t \, q_2}, \frac{1}{(1-s) \, q_2}, \frac{1}{s \, q_2}, \frac{1}{s \, p_1},\\&
 \frac{1}{(1-s) \, p_1}, \frac{1}{s \, p_2}, \frac{1}{(1-s) \, p_2}, \frac{1}{s (1-t)},\\&
 \frac{1}{(1-s) (1-t)}, \frac{1}{s \, t}, \frac{1}{(1-s) \, t}, \frac{1-t}{q_2} \Bigr\} 
 \label{basHatI2}
\end{aligned}\end{equation} 
together with the dual basis 
$|h_i \rangle = {\hat h}_i \, \dd s \wedge \dd t $, where ${\hat h}_i = {\hat e}_i$. 
Upon explicit evaluation, the ${\bf C}$-matrix reads as,
\begin{equation}
    {\bf C} = - \frac{1}{24 \gamma^2} \, D^{-1} \, \hat C \, D^{-1} 
    \, ,
\end{equation}
where
\begin{equation}\begin{small}\begin{aligned}\label{collectDen} 
D = & \, 
\text{diag}( \,y d_5,  y d_4,  d_2,  a  d_6, a-y,  a-y,  
\\ & \qquad \quad 
d_7,  d_7, 1,  1,  1,  1,  a y d_3 )
\end{aligned}\end{small}\end{equation}
and

\begin{equation}
\begin{aligned}
\hat C = \begin{tiny}\left(
\begin{array}{ccccccccccccc}
 -48 & 0 & 0 & 0 & 0 & 0 & 0 & 0 & 0 & 0 & 0 & 0 & 0 \\
 0 & 48 & 0 & 0 & 0 & 0 & 0 & 0 & 0 & 0 & 0 & 0 & 0 \\
 0 & 0 & 32 & 0 & 0 & -8 & 0 & 8 & 0 & 0 & 0 & 0 & 0 \\
 0 & 0 & 0 & 48 & 0 & 0 & 0 & 0 & 0 & 0 & 0 & 0 & 0 \\
 0 & 0 & 0 & 0 & 15 & 5 & -3 & -1 & -3 & -1 & 3 & 1 & 0 \\
 0 & 0 & -8 & 0 & 5 & 7 & -1 & -3 & -1 & -3 & 1 & 3 & 0 \\
 0 & 0 & 0 & 0 & -3 & -1 & 15 & 5 & -3 & -1 & 3 & 1 & 0 \\
 0 & 0 & 8 & 0 & -1 & -3 & 5 & 7 & -1 & -3 & 1 & 3 & 0 \\
 0 & 0 & 0 & 0 & -3 & -1 & -3 & -1 & 3 & 17 & 3 & 1 & 0 \\
 0 & 0 & 0 & 0 & -1 & -3 & -1 & -3 & 17 & 3 & 1 & 3 & 0 \\
 0 & 0 & 0 & 0 & 3 & 1 & 3 & 1 & 3 & 1 & 15 & 5 & 0 \\
 0 & 0 & 0 & 0 & 1 & 3 & 1 & 3 & 1 & 3 & 5 & 15 & 0 \\
 0 & 0 & 0 & 0 & 0 & 0 & 0 & 0 & 0 & 0 & 0 & 0 & -8 \\
\end{array}
\right)\hspace{-2pt} .
\end{tiny}\label{defMatCt}
\end{aligned}\end{equation}

By using the master decomposition formula \eqref{eq:masterdecomposition}, 
$\hat I_1$ is projected onto \eqref{basHatI2} with coefficients
\begin{equation}\begin{aligned}
c_3 = - yz \ , \quad 
c_6 = - yz \frac{a-y}{d_2} \ , \quad 
c_8 =  yz \frac{d_6}{d_2} \ , \quad 
\label{decompHatI1}
\end{aligned}\end{equation}
and $c_i = 0,$ for $i \, \neq \, 3,6,8$.

\textbf{Differential equations.} 
The Pfaffian matrices $\Omega_x$, $x=a,b,y,z$, are too large to be shown here. They are included in the {\it ancillary files}. Strikingly, all four matrices are at most linear in $\gamma$. Furthermore, their $\gamma^0$ part is diagonal and of dLog form. It can be removed by scaling the masters with \eqref{collectDen} upon which we obtain canonical equations for pure functions, in the sense of \cite{hennAlg}.

The master integrals in \eqref{basHatI2} will diverge only if the master forms have poles at the ends of the integration intervals $s,t \eqsp 0,1$. Thus, the first and last master integral, i.e. $J_1$ and $J_{13}$ are finite, in the $\gamma \to 0$ limit,
whereas the Laurent expansion of the master integrals $J_i$ with  
$i \in \{2, \ldots, 8\}$ starts at $1/\gamma$, and that of those for $i\in \{9,10,11,12\}$, at $1/\gamma^2$. Interestingly, the leading logarithmic contribution at the highest pole in $\gamma$ can in all cases be determined by freezing the integration variables $s,t$ at the relevant poles and taking any remaining integral. Uniform transcendentality \cite{lipKot} is manifest in our computations so that these terms are single logarithms at $1/\gamma$ or rational numbers at $1/\gamma^2$. The latter all turn out to be one and source the differential equations. 

Nevertheless, solving up to dilogarithm level only for the master integrals $J_3, \, J_6, \, J_8$ featuring in the decomposition \eqref{decompHatI1}, we merely have to deal with multilinear denominators and $d_8$. The latter is  quadratic only in $a, \, y$. As a consequence, solving the equations in the sequence $z, \, b, \, y, \, a$ we do not run into any involved integrations.

Putting together $\hat I_1$ according to \eqref{decompHatI1}, the $1/\gamma$ part cancels and in the finite part we obtain a single pure function at dilogarithm level with the overall rational factor $y z/d_2$.  Taking into account the contributions from $\tilde I_1$, the total symbol is given by
\begin{align}
\mathcal{S}(I_1)\eqsp\frac{yz}{d_2}s_{2,1,1}
\end{align}
with
\begin{equation}\begin{footnotesize}\begin{aligned} \label{resS3}
s_{2,1,1}  = & \frac{a (1-a) (b-z) y (1-y-z)}{(1-z)} \otimes \frac{(1- a y - z)}{(1-z)}  \\
 - & \frac{a (1-a-b) (b-z) y (1-y)}{(1-b)} \otimes \frac{(1- b - a y)}{(1-b)}  \\
 + & d_2 \otimes \frac{(1-a-b)(1-y)(1-z)}{(1-a)(1-b)(1-y-z)} \\
 + &
d_8 \otimes \frac{(1-a)(1-b - a y)(1-y-z)}{(1-a-b)(1-a y -z)(1-y)} \, .  
\end{aligned}\end{footnotesize}\end{equation}

\section{The three-parameter integral~$I_2$} \label{secIntI3}

Let us now consider the second contribution to $S_2$, namely the integral $I_2$ in \eqref{eq:integrals}. Here, the integral over $r$ and $s$ can be carried out. We are left with the $t$ integration, which can be further split into an elementary part $\tilde I_2$ and a twisted period integral of the same type as $\hat I_1$,
\begin{align}
I_2=\tilde I_2+\hat I_2 \, .
\end{align}
For details on this step, see Appendix~\ref{App:I2}. Here we want to focus entirely on the resulting two-parameter integral
\begin{equation}\begin{aligned}\label{hardCore}
\hat I_2 = \int_{\mathcal{C}_R}
u_2(s,t) \, \varphi_2(s,\,t) = \langle \varphi_2(s,\,t) | \mathcal{C}_R]~~~\\
\text{with }~~u_2(s,t)\eqsp s^\gamma (1-s)^\gamma t^\gamma (1-t)^\gamma p_6^\gamma q_2^\gamma \, ,\\
\varphi_2(s,\,t)=\frac{y^2 z \, (1-t) \, n_4(s,t)}{2 \,p_6(t) \, q_2(s,t)} \, ds \wedge dt \, ,
\end{aligned}\end{equation}
in the $\gamma\rightarrow0$ limit. The intersection theory computation has almost exactly the same features as the former one. 

\noindent 
\textbf{Bases.} The criterion \eqref{eq:leepom2} yields $\nu=12$, therefore $\hat I_2$ can be decomposed in terms of a basis of 12 master integrals, whose corresponding basis of master 2-forms 
$\langle e_i | = {\hat e}_i 
\, \dd s \wedge \dd t $ 
is chosen to be
\begin{equation}\begin{aligned}
 \label{basHat3}
 \{ {\hat e}_i \}_{i=1}^{12} = 
 \Bigl\{& \frac{1-t}{p_6 \, q_2}, \, \frac{1}{q_2}, \, \frac{1-t}{t \, q_2}, \, \frac{1}{(1-s) \, q_2}, \, \frac{1}{s \, q_2}, \\ & \, \frac{1}{s \, p_6},\, \frac{1}{(1-s) \, p_6},
  \, \frac{1}{s (1-t)}, \, \\ &
  \frac{1}{(1-s) (1-t)},   \ \frac{1}{s \, t}, \, \frac{1}{(1-s) \, t}, \, \frac{1-t}{q_2} \Bigr\}  \,, 
\end{aligned}\end{equation}
along with the dual basis 
$|h_i \rangle = {\hat h}_i \, \dd s \wedge \dd t $, with ${\hat h}_i = {\hat e}_i$. In this case, the intersection 
${\bf C}$-matrix reads
\begin{equation}
    {\bf C} = - \frac{1}{20 \gamma^2} \, D^{-1} \, \hat C \, D^{-1} 
    \, ,
\end{equation}
where 
\begin{equation}\begin{small}\begin{aligned}\label{collectDenHat3}
D =\text{diag}( y  d_{10},  y \, d_5,  y  d_4,  a  d_6,  d_2,  d_2,  d_2,  1,  1,  1,  1, a y  d_3 )
\end{aligned}\end{small}\end{equation}
and

\begin{equation}\begin{aligned}
\hat C = \begin{tiny}\left(
\begin{array}{cccccccccccc}
 40 & 0 & 0 & 0 & 0 & 0 & 0 & 0 & 0 & 0 & 0 & 0 \\
 0 & -40 & 0 & 0 & 0 & 0 & 0 & 0 & 0 & 0 & 0 & 0 \\
 0 & 0 & 40 & 0 & 0 & 0 & 0 & 0 & 0 & 0 & 0 & 0 \\
 0 & 0 & 0 & 40 & 0 & 0 & 0 & 0 & 0 & 0 & 0 & 0 \\
 0 & 0 & 0 & 0 & 40 & 0 & 0 & 0 & 0 & 0 & 0 & 0 \\
 0 & 0 & 0 & 0 & 0 & 12 & 4 & 3 & 1 & -3 & -1 & 0 \\
 0 & 0 & 0 & 0 & 0 & 4 & 12 & 1 & 3 & -1 & -3 & 0 \\
 0 & 0 & 0 & 0 & 0 & 3 & 1 & 2 & 14 & 3 & 1 & 0 \\
 0 & 0 & 0 & 0 & 0 & 1 & 3 & 14 & 2 & 1 & 3 & 0 \\
 0 & 0 & 0 & 0 & 0 & -3 & -1 & 3 & 1 & 12 & 4 & 0 \\
 0 & 0 & 0 & 0 & 0 & -1 & -3 & 1 & 3 & 4 & 12 & 0 \\
 0 & 0 & 0 & 0 & 0 & 0 & 0 & 0 & 0 & 0 & 0 & -8 \\
\end{array}
\right)\hspace{-2pt}.
\end{tiny} \label{CtPrime}
\end{aligned}\end{equation}

The projection of $\hat I_2$ onto the basis \eqref{basHat3} is carried out 
by means of the master decomposition formula in \eqref{eq:masterdecomposition}, 
yielding the coefficients
\begin{equation}\label{decompHatI2} 
c_1 = - \frac{y^2 z}{2} \frac{d_{10}}{d_2} \ , \quad 
c_{12} = \frac{y^2 z}{2} \frac{a \, d_3}{d_2} \ , 
\end{equation}
and $c_i = 0,$ for $i = 2,\ldots,11$. As the masters have the denominators listed in \eqref{collectDenHat3}
we find a single pure function with rational factor $y z / d_2$.

\textbf{Differential equations.} 
Taking this rescaling into account we obtain canonical equations. However, the denominators of the Pfaffian matrices $\Omega_x$ (cf. \textit{ancillary files}) contain quite a few non-multilinear polynomials, e.g. $d_5$ as before or the new $d_{10}$. Evaluating only the differential equations for $J_1,\, J_{12}$ to the dilogarithm level solving first in $z$, then in $y, \, b, \, a$ ensures that only elementary integrals are met.

Recall that we had split the original three-parameter Euler integral from \eqref{defS12} into $\hat I_2$ and a directly integrable part $\tilde I_2$. Adding those up, we find
\begin{align}
\mathcal{S}(I_2)\eqsp \frac{yz}{d_2} s_{2,1,2} +  \frac{yz}{d_{11}} s_{2,2,2} 
\end{align}
with
\begin{equation}\begin{footnotesize}\begin{aligned}
s_{2,1,2}  = & d_7 \otimes \frac{(1-b)(1-y-z)}{(1-a-b)(1-z)} 
+ (a-y) \otimes \frac{(1-a)}{(1-y)} \\&
+ d_8 \otimes \frac{(1-a-b)(1 - a y - z)(1-y)}{(1-a)(1-b-a y)(1-y-z)}  \\&
-  \frac{a ( 1-a) (b-z) y (1-y-z)}{(1-z)} \otimes (1 -a y - z) \\&
+ \frac{a (1-a-b) (b-z) y (1-y)}{(1-b)} \otimes (1 - b - a y)  \\&
- (b-z) \otimes \frac{(1-b)}{(1-z)} 
+  a \otimes \frac{(1-a)(1-z)}{(1-a-b)} \\&
+ b \otimes \frac{(1-a)(1-b)}{(1-a-b)} 
- y \otimes \frac{(1-b)(1-y)}{(1-y-z)} \\&
- z \otimes  \frac{(1-y)(1-z)}{(1-y-z)}  
+  \frac{(1-a)}{(1-z)} \otimes (1-y-z) \\&
+ (1-y-z) \otimes (1-a)(1-z)  \\&
-  \frac{(1-y)}{(1-b)} \otimes (1-a-b) \\&
- (1-a-b) \otimes (1-b)(1-y)  , 
\end{aligned}\end{footnotesize}\end{equation}
\begin{equation}\begin{footnotesize}\begin{aligned}
s_{2,2,2}  = &
- d_2 \otimes \frac{(1-a-b)(1-y)(1-z)}{(1-a)(1-b)(1-y-z)}\\&
- d_7 \otimes \frac{(1-b)(1-y-z)}{(1-a-b)(1-z)}
- (a-y) \otimes \frac{(1-a)}{(1-y)} \\&
+ (1-z) \otimes \frac{(1-y-z)}{(1-a)(1-z)}
+ (y z) \otimes \frac{(1-y)(1-z)}{(1-y-z)} \\&
-  (a b) \otimes \frac{(1-a)(1-b)}{(1-a-b)} 
- (1-b) \otimes \frac{(1-a-b)}{(1-b)(1-y)} .
\end{aligned}\end{footnotesize}\end{equation}
Adding in \eqref{resS3}, these formulae show that indeed 
\begin{equation}\begin{aligned}
 S_2 \eqsp I_1 + I_2 \eqsp S_1(a \leftrightarrow b, \, y \leftrightarrow z)
\end{aligned}\end{equation}
up to cuts, as anticipated in \eqref{theFeature}.

For interested readers, we comment on numerical checks of the full solution of the differential equations against the respective integrals in Appendix \ref{sec:num_checks}.

\section{Discussion}

Integrability allows us to construct higher-point correlation functions of gauge invariant composite operators in ${\cal N}=4$ SYM from triangulations \cite{BKV,cushions,shotaThiago}. 
However, field theory results from Feynman graphs can only be recovered when the tiles are glued together again. To date, such gluing processes remain hard to compute. 

In this work, we studied aspects of the simplest generic case, i.e. the two-particle gluing of the five-point function of stress tensor multiplets \cite{shotaThiago2,usFivePoints}. The analytic evaluation of this contribution already requires the evaluation of challenging sums, which, after casting them in the form of (twisted period) Euler-Mellin integrals, we addressed for the first time by means of intersection theory techniques \cite{Mastrolia:2018uzb,fatIntersection2}.

The Euler integrals were decomposed in terms of an independent set of master integrals, generating the corresponding twisted cohomology groups. These computations required the evaluation of intersection numbers for differential 2-forms. The cohomology bases were suitably chosen to obey a canonical system of Pfaffian equations. The part of this system determining the original Euler integrals was then solvable using differential equation techniques.

In order to compute the relevant intersection numbers, we relied on the aforementioned fibration algorithm implemented in \cite{Brunello:2023rpq,Brunello:2024tqf}, but also on an independent {\sc Mathematica} code, purpose written for this new application. The integral decompositions of the present study run on a laptop well within an hour, suggesting that much harder problems can be tackled by these methods
with existing computer resources.

Our first study concerning the application of intersection theory for twisted co-homology to the gluing technique could be further combined with the $D$-module of operators annihilating GKZ functions, along the lines of the Maculay-matrix based algorithm proposed in \cite{Manosh,Chestnov:2023kww} and other algebraic geometry based methods like the algorithms applied in \cite{Brunello:2023fef,Benincasa:2024ptf}. 
We are confident that such an approach could provide an innovative theoretical tool for studying series such as \eqref{defS12} using Pfaffian systems and, in general, prove to be a promising method for evaluating gluing contributions in ${\cal N}=4$ SYM theory.
\section{Acknowledgements}

We are grateful to S. ~Banik and S.~Friot for discussions on the use of the Mellin-Barnes representation for gluing contributions, and W.~Flieger and M.~Mandal for discussions at the University of Padova, as well as with M.~Borinsky, O.~Schnetz and G.~Papanathanasiou at Desy Hamburg. We acknowledge the stimulating scientific environment during the kickoff meeting of the Universe+
ERC synergy grant, held in Leipzig, February 2024, where we discussed the idea of using intersection theory for gluing integrals. 

We would like to additonally thank Vsevolod Chestnov and Giacomo Brunello for independent checks on the intersection numbers computed for this project.

B. Eden holds Heisenberg funding of the Deutsche Forschungsgemeinschaft (DFG), grant Ed 78/7-2 or 441791296, M. Gottwald is supported by the DFG project grant Ed 78/8-1 or 441793388. T. Scherdin’s research is funded by the Deutsche Forschungsgemeinschaft (DFG, German Research Foundation) - Projektnummer 417533893/GRK2575 “Rethinking Quantum Field Theory”. G. Crisanti and P. Mastrolia acknowledge the support of iniziativa specifica INFN {\it Amplitudes}.

\begin{appendix}

\section{Polynomials}\label{App:polynomials}
\noindent
Where helpful we use the shorthands
\begin{equation}\begin{aligned}
B \eqsp 1-b, \, Y \eqsp 1 - y, \, Z \eqsp 1-z, \, T \eqsp 1-t  .
\end{aligned}\end{equation}
\noindent
Polynomials without integration variables:
\begin{equation}\begin{footnotesize}\begin{aligned} \label{defDens}
d_1  = & \, a  b - a b y - y z + a y z \\
d_2  = & \, a b - a b y - a b z - y z + a y z + b y z \\
d_3  = & \, b - b y - z + a z  \\
d_4  = & \, a b - a b y + z - a z - b z  \\
d_5  = & \, \bigl( z \, (4 a b - 8 a^2 b + 4 a^3 b - 4 a b^2 + 4 a^2 b^2 + z - 4 a z \\&\hspace{-8pt}
 + 6 a^2 z - 4 a^3 z + a^4 z - 2 b z + 4 a b z - 2 a^2 b z + b^2 z )\bigr)^{1/2}  \\
d_6  = & \, b - b y - b z + y z - a y z  \\
d_7  = & \, a - y + b y - a z  \\
d_8  = & \, a b - 2 a b y + a b y^2 - a b z - y z + 2 a y z - a^2 y z + b y z  \\
d_9  = & \, a b + y - 2 a y + a^2 y - b y  \\
d_{10}  = & \, B^2 y z^2 - a (B z (b - 3 b y - bz + 3 yz) + a^3 (b-z) y z) \\&\hspace{-8pt}
 - a^2 (b-z) (b - b y - b z + 3 y z - b y z)  \\
d_{11}  = & \, a b - a b z - y z + b y z  
\end{aligned}\end{footnotesize}\end{equation}
Polynomials depending on $t$:
\begin{equation}\begin{footnotesize}\begin{aligned}
p_1  = & \, 1 - a + a \, T - y \, T \, ,  \\
p_2  = & \, -B y \, T  - a Z + B Z + a Z \, T \, ,  \\
p_3  = & \, (a - B - a \, T + a y \, T ) (a y \, T - Z + a Z - a Z \, T) \, ,  \\
p_4  = & \, -y \, T ( a^2 + B - 2 a B - 2 a Z + a^2 Z + B Z \\&  \, -a (a - B + y - 2 a y + B y - Z + a Z) \, T) \, , \\
p_5  = & \, - (1-a)(a-B)y^2 \, T^2 \, ,  \\
p_6  = & \, a (1 - B) Z + (B y z - a B y - a Z + a B Z + a y Z) \, T
\end{aligned}\end{footnotesize}\end{equation}
Polynomials depending on two parameters $s$~and~$t$:
\begin{equation}\begin{footnotesize}\begin{aligned}
n_1 = & \, a  y^2  z \, T  ( - a + B + Z - a Z \\&
+ a (1 - y + Z) \, T - (1 - a + B) y \, s T )\,,\\
n_3  = & \, - (a^2 + B - 2 a B - 2 a Z + a^2 Z + B Z)  \\&
 + (a^2 - a B - a^2 y + B y - a Z + a^2 Z) \, T\\&
 - (1-a)(a-B) y \, s T \, ,  \\
n_4  = & \, (a^2 + B - 2 a B + 2 a Z - a^2 Z - B Z) \\&
 - a (a - B - y + B y + Z - a Z) \, T \, , \\
q_1  = & \, Y - (b \, Y + z) \, t + b z \, t^2  - (a - y + (b y - a z) \, t) \, s \, ,  \\
q_2  = & \, p_3 + p_4 \, s + p_5 \, s^2 \, ,  \\
q_3  = & \, (1 - a + a \, T - y \, s T) (a Z - B Z - a Z \, T + B y \, s T )
\end{aligned}\end{footnotesize}\end{equation}
$\{r, s, t\}$-dependent polynomials, with $R \eqsp 1 - r$:
\begin{equation}\begin{footnotesize}\begin{aligned}
n_2 = & \, a y^2 z \, s T \, (B + Z - a (1 + Z) s \\&
+ (a - y - B y + a Z) s T + (1 - 2 a + B) y R s T) \, , \\
q_4 = & \, q_1^* + q_2^* \, R s T + q_3^* (R s T)^2 \, , \\
q_1^* = & \, (1 - a \, s + (a-y) s T) (-B Z + a Z\, s + (B y - a Z) s T) \, , \\
q_2^* = & \, - y (B - a B - a Z + B Z + a (a - B - Z + a Z) \, s \\&
- (a^2 - a B - a y + 2 B y - a B y - a Z + a^2 Z) s T ) \, , \\
q_3^* = & \, (1-a)(a-B) y^2 \label{eqA5}
\end{aligned}\end{footnotesize}\end{equation}
\section{Rewriting $I_2$}\label{App:I2}
In the following we explain how reduce the three-fold integral $I_2$ in \eqref{eq:integrals} to an integral over a 2-form, similar to $I_1$. Starting with the expression in \eqref{eq:integrals}, we can in fact integrate out first $r$ and then $s$. The total result contains six functions consisting of simple logs with dLog type rational coefficients linear in the variable $t$. This part can be directly integrated and yields two weight 2 pure functions with different rational prefactors. We will not display this simple calculation. The result is of the form $\tilde I_2 = y z (f_1 /d_{11} + f_2/d_2)$ with two pure functions $f_{1,2}$.

Further, two functions $\arctan (q_i(t)/\sqrt{Q(t)})$ are generated, with $Q(t)$ a quadratic polynomial. Up to a sign they come with the same coefficient. By definition
\begin{equation}\label{matchArcTan}
\int \frac{ds}{s^2 + d \, s + c} \eqsp \frac{1}{\sqrt{4 c - d^2}} \, \arctan\left(\frac{d + 2 s}{\sqrt{4 c - d^2}}\right) \, , 
\end{equation}
where we have scaled $s$ so that the coefficient of $s^2$ is unity. This $s$-independent scaling is, of course, secretly present. We now show how this difference of arctan functions may be viewed as a definite integral over the unit interval. Assuming the root denominator $\sqrt{Q}$ to agree with that of the ansatz \eqref{matchArcTan} up to the re-scaling, we first solve
\begin{equation}
d + 2 \cdot 1 \eqsp q_1 \, e \, , \qquad d + 2 \cdot 0 \eqsp q_0 \, e
\end{equation}
finding
\begin{equation}\begin{footnotesize}\begin{aligned}
e^{-1}  = & -(1-a)(a -B) y \, T \, , \\
\frac{d}{e}  = &  a^2 + B - 2 a B - 2 a Z + a^2 Z + B Z \\
& - (a^2 - a B - a y + 2 B y - a B y - a Z + a^2 Z) \, T \, . 
\end{aligned}\end{footnotesize}\end{equation}
Now we should have 
\begin{equation}
4 c - d^2 \eqsp e^2 Q \, . \label{discQ}
\end{equation}
This yields 
\begin{equation}\begin{footnotesize}
\frac{c}{e} =  \frac{( a - a \, T + y \, T-1)(a Z - B Z - a Z \, T + B y \, T)}{y \, T}  .
\end{footnotesize}\end{equation}
Taking into account the prefactor, the hard part of $I_2$ becomes the two parameter integral presented in \eqref{hardCore}. The symbol of the sum $I_2=\hat I_2 + \tilde I_2$ is presented at the end of the calculation in the main text.

\section{Computation of 1-Form Intersection Numbers}\label{sec:1_form_comp}

The intersection number between a $1$-form $\varphi_L$ and a dual $\varphi_R$ (for a given twist $u$) can be computationally defined as:
\begin{equation}\label{eq:int_1_form_comp}
    \langle \varphi_L | \varphi_R \rangle = \sum_{p\in \mathcal{P}_\omega}\Res_{z=p}(\psi_p\, \varphi_R)\,.
\end{equation}
In this expression, $\mathcal{P}_\omega$ is the set of points where the $1$-form $\omega=d\log(u)$ has \textit{poles}. $\psi_p$ is a function defined locally around each pole, implicitly given as the (local) solution to the differential equation
\begin{equation}
    \nabla_\omega \psi_p = \varphi_L\,.
\end{equation}
The solution to this differential equation can be computed by the ansatz
\begin{equation}
\psi_p = \sum_{n=\text{min}}^{\text{max}} c_i (z-p)^n+\mathcal{O}(z-p)^{\text{max}+1}\,.
\end{equation}
Only a finite number of terms in this expansion will contribute to the residue taken in \eqref{eq:int_1_form_comp}, thus this approximate solution to the differential equation system around each pole will yield an exact value for the intersection number.

\section{Computation of 2-Form Intersection Numbers}\label{sec:2_form_comp}

For 2-forms and beyond, the computation of intersection numbers is more involved and there are multiple possible strategies one can take. Here we describe the recursive, or fibration method. 

The core idea of this method is to project any $2$-form intersection number onto a basis of $1$-forms. The result will be a combination of $1$-form intersections, which can once again be computed with a $1$-form algorithm straightforwardly. 

For concreteness let us assume we have $2$-forms $\varphi_L(z_1,z_2)$ and $\varphi_R(z_1,z_2)$ in the variables $z_1$ and $z_2$, as well as a twist $u(z_1,z_2)$. We begin by constructing a basis $\langle e_i^{(1)}|$ and a dual $|h_i^{(1)}\rangle$ with respect to the variable $z_1$. This is done by treating $z_2$ as a constant. The dimensions of this ``inner" basis can be computed with \eqref{eq:leepom2}. Let us call the dimension of this basis $\nu_1$. The $\nu_1\times \nu_1$ $\mathbf{C}$-matrix for this inner basis can then be computed using the $1$-form intersection algorithm given above
\begin{equation}
    \mathbf{C}_{ij}^{(1)}=\langle e_i^{(1)}|h_i^{(1)}\rangle \,.
\end{equation}
The matrix $\mathbf{C}_{ij}^{(1)}$ cannot depend on $z_1$ as it has been integrated out, but in general can (and often will) depend on the other variable $z_2$. We can now decompose the two forms $\varphi_L$ and $\varphi
_R$ onto the basis of $1$-forms above. We have
\begin{equation}
\begin{aligned}
    \langle \varphi_L(z_1,z_2) | &= \sum_{i=1}^{\nu_1} \langle {\varphi_L}_i^{(2)}(z_2) | \wedge \langle {e_i^{(1)}}|\,,\\
    | \varphi_R(z_1,z_2) \rangle &= \sum_{i=1}^{\nu_1} | {\varphi_R}_i^{(2)}(z_2) \rangle \wedge | {e_i^{(1)}}\rangle\,.
\end{aligned}
\end{equation}
The vectors ${\varphi_L}_i^{(2)}(z_2)$ and ${\varphi_R}_i^{(2)}(z_2)$ can be extracted using the master decomposition formula given in \eqref{eq:masterdecomposition}, once again computed by considering $1$-form intersections in $z_1$ and treating $z_2$ as a constant:
\begin{equation}
\begin{aligned}
    \langle {\varphi_L}_i^{(2)}(z_2) &= \langle \varphi_L(z_1,z_2) | h_j \rangle (\mathbf{C^{(1)}})_{ji}^{-1}\\
    | {\varphi_R}_i^{(2)}(z_2) \rangle &= (\mathbf{C^{(1)}})_{ji}^{-1} \langle e_j^{(1)} | \varphi_R \rangle\,.
\end{aligned}
\end{equation}
To finish the intersection computation, we now define a $\nu_1 \times \nu_1$ matrix connection $\Omega^{(2)}$, implicitly defined as
\begin{equation}
    \langle (\partial_{z_2} + \omega_2) e_i^{(1)}| = \sum_{j=1}^{\nu_1} \Omega^{(2)}_{ij} \langle e_j^{(1)}|\,.
\end{equation}
The exact form for $\Omega^{(2)}$ can be computed by once again evaluating $1$-form intersection numbers via projection
\begin{equation}
    \Omega^{(2)}_{ij} = \langle(\partial_{z_2} + \omega_2) e_i^{(1)} | h_k^{(1)} \rangle \mathbf{C}_{kj}^{(1)}\,.
\end{equation}
The 2-form intersection number is finally given by
\begin{equation}\label{eq:intersectT}
    \langle \varphi_L | \varphi_R \rangle = \sum_{p\in \mathcal{P}_\Omega^{(2)}} \left( (\psi_p)_i \, \mathbf{C}_{ij}^{(1)} \, (\varphi_j^{(2)}) \right)\,.
\end{equation}
In this expression, $(\psi_p)_i$ is a \textit{vector} of length $\nu_1$ computed by solving the $\nu_1$ dimensional vector differential equation
\begin{equation}\label{defPsiW}
    (\nabla_{\Omega^{(2)}} (\psi_p))_i = \partial_{z_2} (\psi_p)_i + (\psi_p)_j (\Omega^{(2)})_{ji} =  \varphi^{(2)}_i\,,
\end{equation}
locally around the poles of $\Omega^{(2)}$. Analogously to before, this equation can be solved with a power series ansatz for the vector $(\psi_p)_i$.

\section{Specific Computational Details}\label{sec:comp_details}

We choose $s \eqsp z_1, \, t \eqsp z_2$ for the reduction of both, $\hat I_1, \, \hat I_2$. By using \eqref{eq:leepom2} we find $\nu_1=3$, and choose the inner basis forms to be
\begin{equation}
\{ \hat e_i^{(1)} \} \, = \, \Bigl\{\frac{1}{s \, t}, \frac{1}{(1-s) \, t}, \frac{1-t}{q_2} \Bigr\}\,, \label{basS}
\end{equation}
To handle the roots of $q_2(s,t)$ w.r.t. $s$ we write
\begin{equation}\begin{aligned}
q_2(s,t) \eqsp (s-s^+(t))(s-s^-(t)) \, p_5(t) \, . \label{quadS}
\end{aligned}\end{equation}
In level-1 intersection numbers only integer powers of the discriminant $\mathrm{disc}_s=(s^+-s^-)^2 p_5^2 \,$ appear because of the sum over the expansions around the two poles $s^\mp$. Importantly, while this discriminant is a priori quartic in $t$, it actually factors into $(1-t)^2$ and a new quadratic polynomial whose zeroes $t^\pm$ can be handled in the same manner for the second level intersection numbers.

The level-1 $\mathbf{C}$-matrix is given by
\begin{equation}
\mathbf{C}^{(1)} \, = \, \frac{1}{\gamma} \, D^{-1} \left(\begin{array}{ccc} 3 & 1 & 0 \\ 1 & 3 & 0 \\ 0 & 0 & 2 \end{array}\right) D^{-1} \label{Cs}
\end{equation}
for both $I_1$ and $I_2$. $D$ is the diagonal matrix 
\begin{equation}
D = \text{diag}(2 \, t, \, 2 \, t, \, (1 - a) (1 - a - b) y^2 \, (s^- - s^+) T ) \, . \label{Ds}
\end{equation}

Wherever possible the basis forms should maximally have simple poles --- importantly also at infinity --- in order to minimise the depth of the Taylor expansions necessary in picking residues. The extra factors $t$ and $1-t$ in \eqref{basS} serve this purpose in the later second level computation. 

With the bases \eqref{basHatI2} and \eqref{basHat3}, the second level intersection matrices $\textbf{C}^{(2)}$ can be found from leading order Taylor expansions only, when solving \eqref{defPsiW}. However, the reduction of derivatives of the bases, as required for deriving Pfaffian equations, would in principle necessitate expanding to higher order. This would present a considerable complication in solving for the second level potentials $\psi$ around $t^\mp$. In the following we sketch an order estimate showing that these \emph{root regions} drop out of the computation of $C^{(2)}$, and that the expansion around $t^\mp$ is only needed at leading order in determining the Pfaffian matrices.

Due to the form of $\mathbf{C}^{(1)}$ included in \eqref{eq:masterdecomposition}, the discriminant occurs as a simple pole only in all three coefficients $c_i, \, i \eqsp 1 \ldots 3$, of derivatives of left masters with a denominator factor $q_2$. On the other hand, we do not differentiate the right masters. Yet, including the explicit $C_s$ in the definition of the level 2 intersection numbers \eqref{eq:intersectT} into the equivalent of \eqref{eq:masterdecomposition}, there can be a simple pole in $\mathrm{disc}_s$ in their third coefficient.

We can extract the simple poles of $\Omega^{(2)}$ coming from $t$-linear denominators via the residue theorem. Beyond these, in both reductions $\Omega^{(2)}$ contains a remainder of the form
\begin{equation}\begin{aligned}
R_t \eqsp \left( \begin{array}{ccc} 0 & 0 & 0 \\ 0 & 0 & 0 \\ r_1 & r_2 & r_3 \end{array} \right) \label{defRt}
\end{aligned}\end{equation}
all three entries of which have the denominator $\mathrm{disc}_s$. While $r_1, \, r_2$ have fairly involved numerators, in both cases
\begin{equation}
r_3 \eqsp \left(\gamma - \frac{1}{2}\right) \, \partial_t \log(q_2) \, .
\end{equation}
Since the $t$ intersection number \eqref{eq:intersectT} is defined by a dot product we only need the entry $r_3$. Its expansion in the root regions is simply
\begin{equation}
r_3 \eqsp \left(\gamma - \frac{1}{2}\right) \frac{1}{t-t^\mp}  + \ldots
\end{equation}
from where \eqref{defPsiW} is easy to solve for $\psi_3$ at the leading order. The other two components of the equation merely present consistency conditions which are always satisfied. In conclusion, $\psi_3$ starts at constant order in $t-t^\mp$. Higher contributions are irrelevant because the right forms have only simple poles.

\section{Numerical checks}\label{sec:num_checks}
The integrals $\hat I_1$, $\hat I_2$ are not numerically stable in the physical regimes, i.e. the Euclidean one where the cross ratios are complex and $b \eqsp \bar z, \, a \eqsp \bar y$, or the Minkowskian case in which they are all four real and independent. We therefore perform numerical checks of our solutions of the Pfaffian equations against the integrals in an unphysical regime
\begin{equation}
\{a, \, b, \, y, \, z\} = \{3 + \sqrt{2} \Delta, \, 5 + \sqrt{3} \Delta, 7 + \sqrt{5} \Delta, 11 + \sqrt{7} \Delta \}
\end{equation}
for a range of imaginary increments $\Delta \eqsp i k /4, \, k \eqsp 1,2\ldots 10$ for $I_1$ and $\Delta \eqsp i k /8, \, k \eqsp 1,2\ldots 16$ for $\hat I_2$. 

In integrating the Pfaffian equations at the dilogarithm level we need the simplification 
\begin{equation}
\log( \prod_i x_i ) \eqsp \sum_i \log( x_i ) + i \pi l \label{branches}
\end{equation}
in order to see the independence of the second equation of the first variable etc. For any such replacement it is important to check that all numerical points in our list yield the same $l$. Using these modified replacement rules and introducing integration constants $g_i \eqsp i \pi m_i$ beyond the sum of single logarithms at $1/\gamma$ in the divergent masters $e_i$, as well as constants $h_i$ at the dilogarithm level, we can try to match the general solutions of our differential equations on numerical results for the integrals.

The general solution of the Pfaffian equations for $\hat I_1$ contains only the fixed combination $h_{368} \eqsp h_3 + h_6 - h_8$, while finiteness requires $m_8 \eqsp m_3 + m_6$. Further, $m_{10}$ drops and so does a linear combination of some of the other parameters. 

For a match we equated with a numerical evaluation of $I_1$ with a desired precision of 20 digits. The problem is quite ill-determined. With a bit of hindsight we can see, though, that the equations leave one free parameter as they should. Fixing $h_{368} \eqsp - \zeta(2)$ we find $m_i \, \in \, \{0,-1,\mp 2\}$. The resulting function is appended as a {\sc Mathematica} expression in the {\it ancillary files}. Comparing to the numerical results for the ten points given above and nine similar ones we achieved agreement to 15 digits of precision in all cases. 

Last, the fact that we carefully have to follow \eqref{branches} during its derivation suggests that our solution for $\hat I_1$ is not single-valued. In fact, it apparently has a jump discontinuity in passing from positive to negative increment $k$ for instance, which is not compensated by $\tilde I_1$. Hence a match in this new regime is excluded without changing the integration constants. On the other hand, flipping the sign of $\pi$ --- so that of all $m_i$ --- agreement is recovered.

Finally, we did not try to numerically evaluate the three-parameter integral $I_2$, but rather attempted a fit of the solution of the differential equations for $\hat I_2$ onto numerical data for the same integral. To this end we used the aforementioned 16 points with smaller interval spacing. The system of equations on the integration constants is more well-behaved. A very similar solution exists and the resulting function is also appended. Once again, numerical agreement with $\hat I_2$ is seen up to approximately 15 digits.

\end{appendix}

\bibliographystyle{apsrev4-1} % remove to have titles in the bibliography
\bibliography{cites}% Produces the bibliography via BibTeX.

%merlin.mbs apsrev4-1.bst 2010-07-25 4.21a (PWD, AO, DPC) hacked
%Control: key (0)
%Control: author (72) initials jnrlst
%Control: editor formatted (1) identically to author
%Control: production of article title (-1) disabled
%Control: page (0) single
%Control: year (1) truncated
%Control: production of eprint (0) enabled
\begin{thebibliography}{49}%
\makeatletter
\providecommand \@ifxundefined [1]{%
 \@ifx{#1\undefined}
}%
\providecommand \@ifnum [1]{%
 \ifnum #1\expandafter \@firstoftwo
 \else \expandafter \@secondoftwo
 \fi
}%
\providecommand \@ifx [1]{%
 \ifx #1\expandafter \@firstoftwo
 \else \expandafter \@secondoftwo
 \fi
}%
\providecommand \natexlab [1]{#1}%
\providecommand \enquote  [1]{``#1''}%
\providecommand \bibnamefont  [1]{#1}%
\providecommand \bibfnamefont [1]{#1}%
\providecommand \citenamefont [1]{#1}%
\providecommand \href@noop [0]{\@secondoftwo}%
\providecommand \href [0]{\begingroup \@sanitize@url \@href}%
\providecommand \@href[1]{\@@startlink{#1}\@@href}%
\providecommand \@@href[1]{\endgroup#1\@@endlink}%
\providecommand \@sanitize@url [0]{\catcode `\\12\catcode `\$12\catcode
  `\&12\catcode `\#12\catcode `\^12\catcode `\_12\catcode `\%12\relax}%
\providecommand \@@startlink[1]{}%
\providecommand \@@endlink[0]{}%
\providecommand \url  [0]{\begingroup\@sanitize@url \@url }%
\providecommand \@url [1]{\endgroup\@href {#1}{\urlprefix }}%
\providecommand \urlprefix  [0]{URL }%
\providecommand \Eprint [0]{\href }%
\providecommand \doibase [0]{http://dx.doi.org/}%
\providecommand \selectlanguage [0]{\@gobble}%
\providecommand \bibinfo  [0]{\@secondoftwo}%
\providecommand \bibfield  [0]{\@secondoftwo}%
\providecommand \translation [1]{[#1]}%
\providecommand \BibitemOpen [0]{}%
\providecommand \bibitemStop [0]{}%
\providecommand \bibitemNoStop [0]{.\EOS\space}%
\providecommand \EOS [0]{\spacefactor3000\relax}%
\providecommand \BibitemShut  [1]{\csname bibitem#1\endcsname}%
\let\auto@bib@innerbib\@empty
%</preamble>
\bibitem [{\citenamefont {Grisaru}\ \emph {et~al.}(1980)\citenamefont
  {Grisaru}, \citenamefont {Ro\ifmmode~\check{c}\else \v{c}\fi{}ek},\ and\
  \citenamefont {Siegel}}]{PhysRevLett.45.1063}%
  \BibitemOpen
  \bibfield  {author} {\bibinfo {author} {\bibfnamefont {M.}~\bibnamefont
  {Grisaru}}, \bibinfo {author} {\bibfnamefont {M.}~\bibnamefont
  {Ro\ifmmode~\check{c}\else \v{c}\fi{}ek}}, \ and\ \bibinfo {author}
  {\bibfnamefont {W.}~\bibnamefont {Siegel}},\ }\href {\doibase
  10.1103/PhysRevLett.45.1063} {\bibfield  {journal} {\bibinfo  {journal}
  {Phys. Rev. Lett.}\ }\textbf {\bibinfo {volume} {45}},\ \bibinfo {pages}
  {1063} (\bibinfo {year} {1980})}\BibitemShut {NoStop}%
\bibitem [{\citenamefont {Caswell}\ and\ \citenamefont
  {Zanon}(1981)}]{CASWELL1981152}%
  \BibitemOpen
  \bibfield  {author} {\bibinfo {author} {\bibfnamefont {W.~E.}\ \bibnamefont
  {Caswell}}\ and\ \bibinfo {author} {\bibfnamefont {D.}~\bibnamefont
  {Zanon}},\ }\href {\doibase https://doi.org/10.1016/0370-2693(81)90764-4}
  {\bibfield  {journal} {\bibinfo  {journal} {Physics Letters B}\ }\textbf
  {\bibinfo {volume} {100}},\ \bibinfo {pages} {152} (\bibinfo {year}
  {1981})}\BibitemShut {NoStop}%
\bibitem [{\citenamefont {Mandelstam}(1982)}]{mandelstam1982light}%
  \BibitemOpen
  \bibfield  {author} {\bibinfo {author} {\bibfnamefont {S.}~\bibnamefont
  {Mandelstam}},\ }\href@noop {} {\bibfield  {journal} {\bibinfo  {journal} {Le
  Journal de Physique Colloques}\ }\textbf {\bibinfo {volume} {43}},\ \bibinfo
  {pages} {C3} (\bibinfo {year} {1982})}\BibitemShut {NoStop}%
\bibitem [{\citenamefont {Brink}\ \emph
  {et~al.}(1983{\natexlab{a}})\citenamefont {Brink}, \citenamefont {Lindgren},\
  and\ \citenamefont {{E.W. Nilsson}}}]{BRINK1983401}%
  \BibitemOpen
  \bibfield  {author} {\bibinfo {author} {\bibfnamefont {L.}~\bibnamefont
  {Brink}}, \bibinfo {author} {\bibfnamefont {O.}~\bibnamefont {Lindgren}}, \
  and\ \bibinfo {author} {\bibfnamefont {B.}~\bibnamefont {{E.W. Nilsson}}},\
  }\href {\doibase https://doi.org/10.1016/0550-3213(83)90678-8} {\bibfield
  {journal} {\bibinfo  {journal} {Nuclear Physics B}\ }\textbf {\bibinfo
  {volume} {212}},\ \bibinfo {pages} {401} (\bibinfo {year}
  {1983}{\natexlab{a}})}\BibitemShut {NoStop}%
\bibitem [{\citenamefont {Brink}\ \emph
  {et~al.}(1983{\natexlab{b}})\citenamefont {Brink}, \citenamefont {Lindgren},\
  and\ \citenamefont {{E.W. Nilsson}}}]{BRINK1983323}%
  \BibitemOpen
  \bibfield  {author} {\bibinfo {author} {\bibfnamefont {L.}~\bibnamefont
  {Brink}}, \bibinfo {author} {\bibfnamefont {O.}~\bibnamefont {Lindgren}}, \
  and\ \bibinfo {author} {\bibfnamefont {B.}~\bibnamefont {{E.W. Nilsson}}},\
  }\href {\doibase https://doi.org/10.1016/0370-2693(83)91210-8} {\bibfield
  {journal} {\bibinfo  {journal} {Physics Letters B}\ }\textbf {\bibinfo
  {volume} {123}},\ \bibinfo {pages} {323} (\bibinfo {year}
  {1983}{\natexlab{b}})}\BibitemShut {NoStop}%
\bibitem [{\citenamefont {Howe}\ \emph {et~al.}(1984)\citenamefont {Howe},
  \citenamefont {Stelle},\ and\ \citenamefont {Townsend}}]{HOWE1984125}%
  \BibitemOpen
  \bibfield  {author} {\bibinfo {author} {\bibfnamefont {P.}~\bibnamefont
  {Howe}}, \bibinfo {author} {\bibfnamefont {K.}~\bibnamefont {Stelle}}, \ and\
  \bibinfo {author} {\bibfnamefont {P.}~\bibnamefont {Townsend}},\ }\href
  {\doibase https://doi.org/10.1016/0550-3213(84)90528-5} {\bibfield  {journal}
  {\bibinfo  {journal} {Nuclear Physics B}\ }\textbf {\bibinfo {volume}
  {236}},\ \bibinfo {pages} {125} (\bibinfo {year} {1984})}\BibitemShut
  {NoStop}%
\bibitem [{\citenamefont {Minahan}\ and\ \citenamefont
  {Zarembo}(2003)}]{Minahan:2002ve}%
  \BibitemOpen
  \bibfield  {author} {\bibinfo {author} {\bibfnamefont {J.~A.}\ \bibnamefont
  {Minahan}}\ and\ \bibinfo {author} {\bibfnamefont {K.}~\bibnamefont
  {Zarembo}},\ }\href {\doibase 10.1088/1126-6708/2003/03/013} {\bibfield
  {journal} {\bibinfo  {journal} {JHEP}\ }\textbf {\bibinfo {volume} {03}},\
  \bibinfo {pages} {013} (\bibinfo {year} {2003})},\ \Eprint
  {http://arxiv.org/abs/hep-th/0212208} {arXiv:hep-th/0212208} \BibitemShut
  {NoStop}%
\bibitem [{\citenamefont {Beisert}\ \emph {et~al.}(2004)\citenamefont
  {Beisert}, \citenamefont {Dippel},\ and\ \citenamefont
  {Staudacher}}]{Beisert:2004hm}%
  \BibitemOpen
  \bibfield  {author} {\bibinfo {author} {\bibfnamefont {N.}~\bibnamefont
  {Beisert}}, \bibinfo {author} {\bibfnamefont {V.}~\bibnamefont {Dippel}}, \
  and\ \bibinfo {author} {\bibfnamefont {M.}~\bibnamefont {Staudacher}},\
  }\href {\doibase 10.1088/1126-6708/2004/07/075} {\bibfield  {journal}
  {\bibinfo  {journal} {JHEP}\ }\textbf {\bibinfo {volume} {07}},\ \bibinfo
  {pages} {075} (\bibinfo {year} {2004})},\ \Eprint
  {http://arxiv.org/abs/hep-th/0405001} {arXiv:hep-th/0405001} \BibitemShut
  {NoStop}%
\bibitem [{\citenamefont {Beisert}\ and\ \citenamefont
  {Staudacher}(2005)}]{Beisert:2005fw}%
  \BibitemOpen
  \bibfield  {author} {\bibinfo {author} {\bibfnamefont {N.}~\bibnamefont
  {Beisert}}\ and\ \bibinfo {author} {\bibfnamefont {M.}~\bibnamefont
  {Staudacher}},\ }\href {\doibase 10.1016/j.nuclphysb.2005.06.038} {\bibfield
  {journal} {\bibinfo  {journal} {Nucl. Phys. B}\ }\textbf {\bibinfo {volume}
  {727}},\ \bibinfo {pages} {1} (\bibinfo {year} {2005})},\ \Eprint
  {http://arxiv.org/abs/hep-th/0504190} {arXiv:hep-th/0504190} \BibitemShut
  {NoStop}%
\bibitem [{\citenamefont {Janik}(2006)}]{Janik:2006dc}%
  \BibitemOpen
  \bibfield  {author} {\bibinfo {author} {\bibfnamefont {R.~A.}\ \bibnamefont
  {Janik}},\ }\href {\doibase 10.1103/PhysRevD.73.086006} {\bibfield  {journal}
  {\bibinfo  {journal} {Phys. Rev. D}\ }\textbf {\bibinfo {volume} {73}},\
  \bibinfo {pages} {086006} (\bibinfo {year} {2006})},\ \Eprint
  {http://arxiv.org/abs/hep-th/0603038} {arXiv:hep-th/0603038} \BibitemShut
  {NoStop}%
\bibitem [{\citenamefont {Eden}\ and\ \citenamefont {Staudacher}(2006)}]{BES1}%
  \BibitemOpen
  \bibfield  {author} {\bibinfo {author} {\bibfnamefont {B.}~\bibnamefont
  {Eden}}\ and\ \bibinfo {author} {\bibfnamefont {M.}~\bibnamefont
  {Staudacher}},\ }\href {\doibase 10.1088/1742-5468/2006/11/P11014} {\bibfield
   {journal} {\bibinfo  {journal} {J. Stat. Mech.}\ }\textbf {\bibinfo {volume}
  {0611}},\ \bibinfo {pages} {P11014} (\bibinfo {year} {2006})},\ \Eprint
  {http://arxiv.org/abs/hep-th/0603157} {arXiv:hep-th/0603157} \BibitemShut
  {NoStop}%
\bibitem [{\citenamefont {Beisert}\ \emph {et~al.}(2007)\citenamefont
  {Beisert}, \citenamefont {Eden},\ and\ \citenamefont {Staudacher}}]{BES2}%
  \BibitemOpen
  \bibfield  {author} {\bibinfo {author} {\bibfnamefont {N.}~\bibnamefont
  {Beisert}}, \bibinfo {author} {\bibfnamefont {B.}~\bibnamefont {Eden}}, \
  and\ \bibinfo {author} {\bibfnamefont {M.}~\bibnamefont {Staudacher}},\
  }\href {\doibase 10.1088/1742-5468/2007/01/P01021} {\bibfield  {journal}
  {\bibinfo  {journal} {J. Stat. Mech.}\ }\textbf {\bibinfo {volume} {0701}},\
  \bibinfo {pages} {P01021} (\bibinfo {year} {2007})},\ \Eprint
  {http://arxiv.org/abs/hep-th/0610251} {arXiv:hep-th/0610251} \BibitemShut
  {NoStop}%
\bibitem [{\citenamefont {Benna}\ \emph {et~al.}(2007)\citenamefont {Benna},
  \citenamefont {Benvenuti}, \citenamefont {Klebanov},\ and\ \citenamefont
  {Scardicchio}}]{Benna:2006nd}%
  \BibitemOpen
  \bibfield  {author} {\bibinfo {author} {\bibfnamefont {M.~K.}\ \bibnamefont
  {Benna}}, \bibinfo {author} {\bibfnamefont {S.}~\bibnamefont {Benvenuti}},
  \bibinfo {author} {\bibfnamefont {I.~R.}\ \bibnamefont {Klebanov}}, \ and\
  \bibinfo {author} {\bibfnamefont {A.}~\bibnamefont {Scardicchio}},\ }\href
  {\doibase 10.1103/PhysRevLett.98.131603} {\bibfield  {journal} {\bibinfo
  {journal} {Phys. Rev. Lett.}\ }\textbf {\bibinfo {volume} {98}},\ \bibinfo
  {pages} {131603} (\bibinfo {year} {2007})},\ \Eprint
  {http://arxiv.org/abs/hep-th/0611135} {arXiv:hep-th/0611135} \BibitemShut
  {NoStop}%
\bibitem [{\citenamefont {Basso}\ \emph {et~al.}(2013)\citenamefont {Basso},
  \citenamefont {Sever},\ and\ \citenamefont {Vieira}}]{Basso:2013vsa}%
  \BibitemOpen
  \bibfield  {author} {\bibinfo {author} {\bibfnamefont {B.}~\bibnamefont
  {Basso}}, \bibinfo {author} {\bibfnamefont {A.}~\bibnamefont {Sever}}, \ and\
  \bibinfo {author} {\bibfnamefont {P.}~\bibnamefont {Vieira}},\ }\href
  {\doibase 10.1103/PhysRevLett.111.091602} {\bibfield  {journal} {\bibinfo
  {journal} {Phys. Rev. Lett.}\ }\textbf {\bibinfo {volume} {111}},\ \bibinfo
  {pages} {091602} (\bibinfo {year} {2013})},\ \Eprint
  {http://arxiv.org/abs/1303.1396} {arXiv:1303.1396 [hep-th]} \BibitemShut
  {NoStop}%
\bibitem [{\citenamefont {Basso}\ \emph
  {et~al.}(2014{\natexlab{a}})\citenamefont {Basso}, \citenamefont {Sever},\
  and\ \citenamefont {Vieira}}]{Basso:2014koa}%
  \BibitemOpen
  \bibfield  {author} {\bibinfo {author} {\bibfnamefont {B.}~\bibnamefont
  {Basso}}, \bibinfo {author} {\bibfnamefont {A.}~\bibnamefont {Sever}}, \ and\
  \bibinfo {author} {\bibfnamefont {P.}~\bibnamefont {Vieira}},\ }\href
  {\doibase 10.1007/JHEP08(2014)085} {\bibfield  {journal} {\bibinfo  {journal}
  {JHEP}\ }\textbf {\bibinfo {volume} {08}},\ \bibinfo {pages} {085} (\bibinfo
  {year} {2014}{\natexlab{a}})},\ \Eprint {http://arxiv.org/abs/1402.3307}
  {arXiv:1402.3307 [hep-th]} \BibitemShut {NoStop}%
\bibitem [{\citenamefont {Basso}\ \emph
  {et~al.}(2014{\natexlab{b}})\citenamefont {Basso}, \citenamefont {Sever},\
  and\ \citenamefont {Vieira}}]{Basso:2014nra}%
  \BibitemOpen
  \bibfield  {author} {\bibinfo {author} {\bibfnamefont {B.}~\bibnamefont
  {Basso}}, \bibinfo {author} {\bibfnamefont {A.}~\bibnamefont {Sever}}, \ and\
  \bibinfo {author} {\bibfnamefont {P.}~\bibnamefont {Vieira}},\ }\href
  {\doibase 10.1007/JHEP09(2014)149} {\bibfield  {journal} {\bibinfo  {journal}
  {JHEP}\ }\textbf {\bibinfo {volume} {09}},\ \bibinfo {pages} {149} (\bibinfo
  {year} {2014}{\natexlab{b}})},\ \Eprint {http://arxiv.org/abs/1407.1736}
  {arXiv:1407.1736 [hep-th]} \BibitemShut {NoStop}%
\bibitem [{\citenamefont {Papathanasiou}(2013)}]{Papathanasiou:2013uoa}%
  \BibitemOpen
  \bibfield  {author} {\bibinfo {author} {\bibfnamefont {G.}~\bibnamefont
  {Papathanasiou}},\ }\href {\doibase 10.1007/JHEP11(2013)150} {\bibfield
  {journal} {\bibinfo  {journal} {JHEP}\ }\textbf {\bibinfo {volume} {11}},\
  \bibinfo {pages} {150} (\bibinfo {year} {2013})},\ \Eprint
  {http://arxiv.org/abs/1310.5735} {arXiv:1310.5735 [hep-th]} \BibitemShut
  {NoStop}%
\bibitem [{\citenamefont {C\'ordova}(2017)}]{Cordova:2016woh}%
  \BibitemOpen
  \bibfield  {author} {\bibinfo {author} {\bibfnamefont {L.}~\bibnamefont
  {C\'ordova}},\ }\href {\doibase 10.1007/JHEP01(2017)051} {\bibfield
  {journal} {\bibinfo  {journal} {JHEP}\ }\textbf {\bibinfo {volume} {01}},\
  \bibinfo {pages} {051} (\bibinfo {year} {2017})},\ \Eprint
  {http://arxiv.org/abs/1606.00423} {arXiv:1606.00423 [hep-th]} \BibitemShut
  {NoStop}%
\bibitem [{\citenamefont {Lam}\ and\ \citenamefont {von
  Hippel}(2016)}]{Lam:2016rel}%
  \BibitemOpen
  \bibfield  {author} {\bibinfo {author} {\bibfnamefont {H.~T.}\ \bibnamefont
  {Lam}}\ and\ \bibinfo {author} {\bibfnamefont {M.}~\bibnamefont {von
  Hippel}},\ }\href {\doibase 10.1007/JHEP12(2016)011} {\bibfield  {journal}
  {\bibinfo  {journal} {JHEP}\ }\textbf {\bibinfo {volume} {12}},\ \bibinfo
  {pages} {011} (\bibinfo {year} {2016})},\ \Eprint
  {http://arxiv.org/abs/1608.08116} {arXiv:1608.08116 [hep-th]} \BibitemShut
  {NoStop}%
\bibitem [{\citenamefont {Basso}\ \emph {et~al.}(2015)\citenamefont {Basso},
  \citenamefont {Komatsu},\ and\ \citenamefont {Vieira}}]{BKV}%
  \BibitemOpen
  \bibfield  {author} {\bibinfo {author} {\bibfnamefont {B.}~\bibnamefont
  {Basso}}, \bibinfo {author} {\bibfnamefont {S.}~\bibnamefont {Komatsu}}, \
  and\ \bibinfo {author} {\bibfnamefont {P.}~\bibnamefont {Vieira}},\
  }\href@noop {} {\  (\bibinfo {year} {2015})},\ \Eprint
  {http://arxiv.org/abs/1505.06745} {arXiv:1505.06745 [hep-th]} \BibitemShut
  {NoStop}%
\bibitem [{\citenamefont {Eden}\ and\ \citenamefont
  {Sfondrini}(2017)}]{cushions}%
  \BibitemOpen
  \bibfield  {author} {\bibinfo {author} {\bibfnamefont {B.}~\bibnamefont
  {Eden}}\ and\ \bibinfo {author} {\bibfnamefont {A.}~\bibnamefont
  {Sfondrini}},\ }\href {\doibase 10.1007/JHEP10(2017)098} {\bibfield
  {journal} {\bibinfo  {journal} {JHEP}\ }\textbf {\bibinfo {volume} {10}},\
  \bibinfo {pages} {098} (\bibinfo {year} {2017})},\ \Eprint
  {http://arxiv.org/abs/1611.05436} {arXiv:1611.05436 [hep-th]} \BibitemShut
  {NoStop}%
\bibitem [{\citenamefont {Fleury}\ and\ \citenamefont
  {Komatsu}(2017)}]{shotaThiago}%
  \BibitemOpen
  \bibfield  {author} {\bibinfo {author} {\bibfnamefont {T.}~\bibnamefont
  {Fleury}}\ and\ \bibinfo {author} {\bibfnamefont {S.}~\bibnamefont
  {Komatsu}},\ }\href {\doibase 10.1007/JHEP01(2017)130} {\bibfield  {journal}
  {\bibinfo  {journal} {JHEP}\ }\textbf {\bibinfo {volume} {01}},\ \bibinfo
  {pages} {130} (\bibinfo {year} {2017})},\ \Eprint
  {http://arxiv.org/abs/1611.05577} {arXiv:1611.05577 [hep-th]} \BibitemShut
  {NoStop}%
\bibitem [{\citenamefont {Bargheer}\ \emph
  {et~al.}(2019{\natexlab{a}})\citenamefont {Bargheer}, \citenamefont
  {Coronado},\ and\ \citenamefont {Vieira}}]{Bargheer:2019exp}%
  \BibitemOpen
  \bibfield  {author} {\bibinfo {author} {\bibfnamefont {T.}~\bibnamefont
  {Bargheer}}, \bibinfo {author} {\bibfnamefont {F.}~\bibnamefont {Coronado}},
  \ and\ \bibinfo {author} {\bibfnamefont {P.}~\bibnamefont {Vieira}},\
  }\href@noop {} {\  (\bibinfo {year} {2019}{\natexlab{a}})},\ \Eprint
  {http://arxiv.org/abs/1909.04077} {arXiv:1909.04077 [hep-th]} \BibitemShut
  {NoStop}%
\bibitem [{\citenamefont {Bargheer}\ \emph
  {et~al.}(2019{\natexlab{b}})\citenamefont {Bargheer}, \citenamefont
  {Coronado},\ and\ \citenamefont {Vieira}}]{Bargheer:2019kxb}%
  \BibitemOpen
  \bibfield  {author} {\bibinfo {author} {\bibfnamefont {T.}~\bibnamefont
  {Bargheer}}, \bibinfo {author} {\bibfnamefont {F.}~\bibnamefont {Coronado}},
  \ and\ \bibinfo {author} {\bibfnamefont {P.}~\bibnamefont {Vieira}},\ }\href
  {\doibase 10.1007/JHEP08(2019)162} {\bibfield  {journal} {\bibinfo  {journal}
  {JHEP}\ }\textbf {\bibinfo {volume} {08}},\ \bibinfo {pages} {162} (\bibinfo
  {year} {2019}{\natexlab{b}})},\ \Eprint {http://arxiv.org/abs/1904.00965}
  {arXiv:1904.00965 [hep-th]} \BibitemShut {NoStop}%
\bibitem [{\citenamefont {Kostov}\ \emph {et~al.}(2019)\citenamefont {Kostov},
  \citenamefont {Petkova},\ and\ \citenamefont {Serban}}]{Kostov:2019stn}%
  \BibitemOpen
  \bibfield  {author} {\bibinfo {author} {\bibfnamefont {I.}~\bibnamefont
  {Kostov}}, \bibinfo {author} {\bibfnamefont {V.~B.}\ \bibnamefont {Petkova}},
  \ and\ \bibinfo {author} {\bibfnamefont {D.}~\bibnamefont {Serban}},\ }\href
  {\doibase 10.1103/PhysRevLett.122.231601} {\bibfield  {journal} {\bibinfo
  {journal} {Phys. Rev. Lett.}\ }\textbf {\bibinfo {volume} {122}},\ \bibinfo
  {pages} {231601} (\bibinfo {year} {2019})},\ \Eprint
  {http://arxiv.org/abs/1903.05038} {arXiv:1903.05038 [hep-th]} \BibitemShut
  {NoStop}%
\bibitem [{\citenamefont {Arutyunov}\ \emph {et~al.}(2009)\citenamefont
  {Arutyunov}, \citenamefont {de~Leeuw},\ and\ \citenamefont
  {Torrielli}}]{glebBound}%
  \BibitemOpen
  \bibfield  {author} {\bibinfo {author} {\bibfnamefont {G.}~\bibnamefont
  {Arutyunov}}, \bibinfo {author} {\bibfnamefont {M.}~\bibnamefont {de~Leeuw}},
  \ and\ \bibinfo {author} {\bibfnamefont {A.}~\bibnamefont {Torrielli}},\
  }\href {\doibase 10.1016/j.nuclphysb.2009.03.024} {\bibfield  {journal}
  {\bibinfo  {journal} {Nucl. Phys. B}\ }\textbf {\bibinfo {volume} {819}},\
  \bibinfo {pages} {319} (\bibinfo {year} {2009})},\ \Eprint
  {http://arxiv.org/abs/0902.0183} {arXiv:0902.0183 [hep-th]} \BibitemShut
  {NoStop}%
\bibitem [{\citenamefont {Arutyunov}\ and\ \citenamefont
  {Frolov}(2009)}]{glebSergey}%
  \BibitemOpen
  \bibfield  {author} {\bibinfo {author} {\bibfnamefont {G.}~\bibnamefont
  {Arutyunov}}\ and\ \bibinfo {author} {\bibfnamefont {S.}~\bibnamefont
  {Frolov}},\ }\href {\doibase 10.1088/1751-8113/42/42/425401} {\bibfield
  {journal} {\bibinfo  {journal} {J. Phys. A}\ }\textbf {\bibinfo {volume}
  {42}},\ \bibinfo {pages} {425401} (\bibinfo {year} {2009})},\ \Eprint
  {http://arxiv.org/abs/0904.4575} {arXiv:0904.4575 [hep-th]} \BibitemShut
  {NoStop}%
\bibitem [{\citenamefont {Fleury}\ and\ \citenamefont
  {Komatsu}(2018)}]{shotaThiago2}%
  \BibitemOpen
  \bibfield  {author} {\bibinfo {author} {\bibfnamefont {T.}~\bibnamefont
  {Fleury}}\ and\ \bibinfo {author} {\bibfnamefont {S.}~\bibnamefont
  {Komatsu}},\ }\href {\doibase 10.1007/JHEP02(2018)177} {\bibfield  {journal}
  {\bibinfo  {journal} {JHEP}\ }\textbf {\bibinfo {volume} {02}},\ \bibinfo
  {pages} {177} (\bibinfo {year} {2018})},\ \Eprint
  {http://arxiv.org/abs/1711.05327} {arXiv:1711.05327 [hep-th]} \BibitemShut
  {NoStop}%
\bibitem [{\citenamefont {De~Leeuw}\ \emph
  {et~al.}(2020{\natexlab{a}})\citenamefont {De~Leeuw}, \citenamefont {Eden},
  \citenamefont {Le~Plat}, \citenamefont {Meier},\ and\ \citenamefont
  {Sfondrini}}]{usFivePoints}%
  \BibitemOpen
  \bibfield  {author} {\bibinfo {author} {\bibfnamefont {M.}~\bibnamefont
  {De~Leeuw}}, \bibinfo {author} {\bibfnamefont {B.}~\bibnamefont {Eden}},
  \bibinfo {author} {\bibfnamefont {D.}~\bibnamefont {Le~Plat}}, \bibinfo
  {author} {\bibfnamefont {T.}~\bibnamefont {Meier}}, \ and\ \bibinfo {author}
  {\bibfnamefont {A.}~\bibnamefont {Sfondrini}},\ }\href {\doibase
  10.1007/JHEP09(2020)039} {\bibfield  {journal} {\bibinfo  {journal} {JHEP}\
  }\textbf {\bibinfo {volume} {09}},\ \bibinfo {pages} {039} (\bibinfo {year}
  {2020}{\natexlab{a}})},\ \Eprint {http://arxiv.org/abs/1912.12231}
  {arXiv:1912.12231 [hep-th]} \BibitemShut {NoStop}%
\bibitem [{\citenamefont {Aomoto}\ and\ \citenamefont
  {Kita}(2011)}]{interTheo1}%
  \BibitemOpen
  \bibfield  {author} {\bibinfo {author} {\bibfnamefont {K.}~\bibnamefont
  {Aomoto}}\ and\ \bibinfo {author} {\bibfnamefont {M.}~\bibnamefont {Kita}},\
  }\href {\doibase 10.1007/978-4-431-53938-4} {\emph {\bibinfo {title} {Theory
  of Hypergeometric Functions}}}\ (\bibinfo  {publisher} {Springer Japan},\
  \bibinfo {year} {2011})\BibitemShut {NoStop}%
\bibitem [{\citenamefont {Mastrolia}\ and\ \citenamefont
  {Mizera}(2019)}]{Mastrolia:2018uzb}%
  \BibitemOpen
  \bibfield  {author} {\bibinfo {author} {\bibfnamefont {P.}~\bibnamefont
  {Mastrolia}}\ and\ \bibinfo {author} {\bibfnamefont {S.}~\bibnamefont
  {Mizera}},\ }\href {\doibase 10.1007/JHEP02(2019)139} {\bibfield  {journal}
  {\bibinfo  {journal} {JHEP}\ }\textbf {\bibinfo {volume} {02}},\ \bibinfo
  {pages} {139} (\bibinfo {year} {2019})},\ \Eprint
  {http://arxiv.org/abs/1810.03818} {arXiv:1810.03818 [hep-th]} \BibitemShut
  {NoStop}%
\bibitem [{\citenamefont {Frellesvig}\ \emph {et~al.}(2021)\citenamefont
  {Frellesvig}, \citenamefont {Gasparotto}, \citenamefont {Laporta},
  \citenamefont {Mandal}, \citenamefont {Mastrolia}, \citenamefont
  {Mattiazzi},\ and\ \citenamefont {Mizera}}]{fatIntersection2}%
  \BibitemOpen
  \bibfield  {author} {\bibinfo {author} {\bibfnamefont {H.}~\bibnamefont
  {Frellesvig}}, \bibinfo {author} {\bibfnamefont {F.}~\bibnamefont
  {Gasparotto}}, \bibinfo {author} {\bibfnamefont {S.}~\bibnamefont {Laporta}},
  \bibinfo {author} {\bibfnamefont {M.~K.}\ \bibnamefont {Mandal}}, \bibinfo
  {author} {\bibfnamefont {P.}~\bibnamefont {Mastrolia}}, \bibinfo {author}
  {\bibfnamefont {L.}~\bibnamefont {Mattiazzi}}, \ and\ \bibinfo {author}
  {\bibfnamefont {S.}~\bibnamefont {Mizera}},\ }\href {\doibase
  10.1007/JHEP03(2021)027} {\bibfield  {journal} {\bibinfo  {journal} {JHEP}\
  }\textbf {\bibinfo {volume} {03}},\ \bibinfo {pages} {027} (\bibinfo {year}
  {2021})},\ \Eprint {http://arxiv.org/abs/2008.04823} {arXiv:2008.04823
  [hep-th]} \BibitemShut {NoStop}%
\bibitem [{\citenamefont {De~Leeuw}\ \emph
  {et~al.}(2020{\natexlab{b}})\citenamefont {De~Leeuw}, \citenamefont {Eden},\
  and\ \citenamefont {Sfondrini}}]{DeLeeuw:2020ifb}%
  \BibitemOpen
  \bibfield  {author} {\bibinfo {author} {\bibfnamefont {M.}~\bibnamefont
  {De~Leeuw}}, \bibinfo {author} {\bibfnamefont {B.}~\bibnamefont {Eden}}, \
  and\ \bibinfo {author} {\bibfnamefont {A.}~\bibnamefont {Sfondrini}},\ }\href
  {\doibase 10.1103/PhysRevD.102.126001} {\bibfield  {journal} {\bibinfo
  {journal} {Phys. Rev. D}\ }\textbf {\bibinfo {volume} {102}},\ \bibinfo
  {pages} {126001} (\bibinfo {year} {2020}{\natexlab{b}})},\ \Eprint
  {http://arxiv.org/abs/2008.01378} {arXiv:2008.01378 [hep-th]} \BibitemShut
  {NoStop}%
\bibitem [{\citenamefont {Goncharov}(1998)}]{Goncharov:1998kja}%
  \BibitemOpen
  \bibfield  {author} {\bibinfo {author} {\bibfnamefont {A.~B.}\ \bibnamefont
  {Goncharov}},\ }\href {\doibase 10.4310/MRL.1998.v5.n4.a7} {\bibfield
  {journal} {\bibinfo  {journal} {Math. Res. Lett.}\ }\textbf {\bibinfo
  {volume} {5}},\ \bibinfo {pages} {497} (\bibinfo {year} {1998})},\ \Eprint
  {http://arxiv.org/abs/1105.2076} {arXiv:1105.2076 [math.AG]} \BibitemShut
  {NoStop}%
\bibitem [{\citenamefont {Goncharov}\ \emph {et~al.}(2010)\citenamefont
  {Goncharov}, \citenamefont {Spradlin}, \citenamefont {Vergu},\ and\
  \citenamefont {Volovich}}]{Goncharov:2010jf}%
  \BibitemOpen
  \bibfield  {author} {\bibinfo {author} {\bibfnamefont {A.~B.}\ \bibnamefont
  {Goncharov}}, \bibinfo {author} {\bibfnamefont {M.}~\bibnamefont {Spradlin}},
  \bibinfo {author} {\bibfnamefont {C.}~\bibnamefont {Vergu}}, \ and\ \bibinfo
  {author} {\bibfnamefont {A.}~\bibnamefont {Volovich}},\ }\href {\doibase
  10.1103/PhysRevLett.105.151605} {\bibfield  {journal} {\bibinfo  {journal}
  {Phys. Rev. Lett.}\ }\textbf {\bibinfo {volume} {105}},\ \bibinfo {pages}
  {151605} (\bibinfo {year} {2010})},\ \Eprint {http://arxiv.org/abs/1006.5703}
  {arXiv:1006.5703 [hep-th]} \BibitemShut {NoStop}%
\bibitem [{\citenamefont {Lee}\ and\ \citenamefont
  {Pomeransky}(2013)}]{Lee:2013hzt}%
  \BibitemOpen
  \bibfield  {author} {\bibinfo {author} {\bibfnamefont {R.~N.}\ \bibnamefont
  {Lee}}\ and\ \bibinfo {author} {\bibfnamefont {A.~A.}\ \bibnamefont
  {Pomeransky}},\ }\href {\doibase 10.1007/JHEP11(2013)165} {\bibfield
  {journal} {\bibinfo  {journal} {JHEP}\ }\textbf {\bibinfo {volume} {11}},\
  \bibinfo {pages} {165} (\bibinfo {year} {2013})},\ \Eprint
  {http://arxiv.org/abs/1308.6676} {arXiv:1308.6676 [hep-ph]} \BibitemShut
  {NoStop}%
\bibitem [{\citenamefont {Cho}\ and\ \citenamefont
  {Matsumoto}(1995)}]{cho1995}%
  \BibitemOpen
  \bibfield  {author} {\bibinfo {author} {\bibfnamefont {K.}~\bibnamefont
  {Cho}}\ and\ \bibinfo {author} {\bibfnamefont {K.}~\bibnamefont
  {Matsumoto}},\ }\href {\doibase 10.1017/S0027763000005304} {\bibfield
  {journal} {\bibinfo  {journal} {Nagoya Math. J.}\ }\textbf {\bibinfo {volume}
  {139}},\ \bibinfo {pages} {67} (\bibinfo {year} {1995})}\BibitemShut
  {NoStop}%
\bibitem [{\citenamefont {Matsumoto}(1998)}]{matsumoto_1998}%
  \BibitemOpen
  \bibfield  {author} {\bibinfo {author} {\bibfnamefont {K.}~\bibnamefont
  {Matsumoto}},\ }\href@noop {} {\bibfield  {journal} {\bibinfo  {journal}
  {Osaka Journal of Mathematics}\ }\textbf {\bibinfo {volume} {35}},\ \bibinfo
  {pages} {873 } (\bibinfo {year} {1998})}\BibitemShut {NoStop}%
\bibitem [{\citenamefont {Frellesvig}\ \emph
  {et~al.}(2019{\natexlab{a}})\citenamefont {Frellesvig}, \citenamefont
  {Gasparotto}, \citenamefont {Laporta}, \citenamefont {Mandal}, \citenamefont
  {Mastrolia}, \citenamefont {Mattiazzi},\ and\ \citenamefont
  {Mizera}}]{Frellesvig:2019kgj}%
  \BibitemOpen
  \bibfield  {author} {\bibinfo {author} {\bibfnamefont {H.}~\bibnamefont
  {Frellesvig}}, \bibinfo {author} {\bibfnamefont {F.}~\bibnamefont
  {Gasparotto}}, \bibinfo {author} {\bibfnamefont {S.}~\bibnamefont {Laporta}},
  \bibinfo {author} {\bibfnamefont {M.~K.}\ \bibnamefont {Mandal}}, \bibinfo
  {author} {\bibfnamefont {P.}~\bibnamefont {Mastrolia}}, \bibinfo {author}
  {\bibfnamefont {L.}~\bibnamefont {Mattiazzi}}, \ and\ \bibinfo {author}
  {\bibfnamefont {S.}~\bibnamefont {Mizera}},\ }\href {\doibase
  10.1007/JHEP05(2019)153} {\bibfield  {journal} {\bibinfo  {journal} {JHEP}\
  }\textbf {\bibinfo {volume} {05}},\ \bibinfo {pages} {153} (\bibinfo {year}
  {2019}{\natexlab{a}})},\ \Eprint {http://arxiv.org/abs/1901.11510}
  {arXiv:1901.11510 [hep-ph]} \BibitemShut {NoStop}%
\bibitem [{\citenamefont {Mizera}(2020)}]{Mizera:2019gea}%
  \BibitemOpen
  \bibfield  {author} {\bibinfo {author} {\bibfnamefont {S.}~\bibnamefont
  {Mizera}},\ }\emph {\bibinfo {title} {{Aspects of Scattering Amplitudes and
  Moduli Space Localization}}},\ \href {\doibase 10.1007/978-3-030-53010-5}
  {Ph.D. thesis},\ \bibinfo  {school} {Princeton, Inst. Advanced Study}
  (\bibinfo {year} {2020}),\ \Eprint {http://arxiv.org/abs/1906.02099}
  {arXiv:1906.02099 [hep-th]} \BibitemShut {NoStop}%
\bibitem [{\citenamefont {Frellesvig}\ \emph
  {et~al.}(2019{\natexlab{b}})\citenamefont {Frellesvig}, \citenamefont
  {Gasparotto}, \citenamefont {Mandal}, \citenamefont {Mastrolia},
  \citenamefont {Mattiazzi},\ and\ \citenamefont
  {Mizera}}]{Frellesvig:2019uqt}%
  \BibitemOpen
  \bibfield  {author} {\bibinfo {author} {\bibfnamefont {H.}~\bibnamefont
  {Frellesvig}}, \bibinfo {author} {\bibfnamefont {F.}~\bibnamefont
  {Gasparotto}}, \bibinfo {author} {\bibfnamefont {M.~K.}\ \bibnamefont
  {Mandal}}, \bibinfo {author} {\bibfnamefont {P.}~\bibnamefont {Mastrolia}},
  \bibinfo {author} {\bibfnamefont {L.}~\bibnamefont {Mattiazzi}}, \ and\
  \bibinfo {author} {\bibfnamefont {S.}~\bibnamefont {Mizera}},\ }\href
  {\doibase 10.1103/PhysRevLett.123.201602} {\bibfield  {journal} {\bibinfo
  {journal} {Phys. Rev. Lett.}\ }\textbf {\bibinfo {volume} {123}},\ \bibinfo
  {pages} {201602} (\bibinfo {year} {2019}{\natexlab{b}})},\ \Eprint
  {http://arxiv.org/abs/1907.02000} {arXiv:1907.02000 [hep-th]} \BibitemShut
  {NoStop}%
\bibitem [{\citenamefont {Brunello}\ \emph
  {et~al.}(2024{\natexlab{a}})\citenamefont {Brunello}, \citenamefont
  {Chestnov}, \citenamefont {Crisanti}, \citenamefont {Frellesvig},
  \citenamefont {Mandal},\ and\ \citenamefont {Mastrolia}}]{Brunello:2023rpq}%
  \BibitemOpen
  \bibfield  {author} {\bibinfo {author} {\bibfnamefont {G.}~\bibnamefont
  {Brunello}}, \bibinfo {author} {\bibfnamefont {V.}~\bibnamefont {Chestnov}},
  \bibinfo {author} {\bibfnamefont {G.}~\bibnamefont {Crisanti}}, \bibinfo
  {author} {\bibfnamefont {H.}~\bibnamefont {Frellesvig}}, \bibinfo {author}
  {\bibfnamefont {M.~K.}\ \bibnamefont {Mandal}}, \ and\ \bibinfo {author}
  {\bibfnamefont {P.}~\bibnamefont {Mastrolia}},\ }\href {\doibase
  10.1007/JHEP09(2024)015} {\bibfield  {journal} {\bibinfo  {journal} {JHEP}\
  }\textbf {\bibinfo {volume} {09}},\ \bibinfo {pages} {015} (\bibinfo {year}
  {2024}{\natexlab{a}})},\ \Eprint {http://arxiv.org/abs/2401.01897}
  {arXiv:2401.01897 [hep-th]} \BibitemShut {NoStop}%
\bibitem [{\citenamefont {Brunello}\ \emph
  {et~al.}(2024{\natexlab{b}})\citenamefont {Brunello}, \citenamefont
  {Chestnov},\ and\ \citenamefont {Mastrolia}}]{Brunello:2024tqf}%
  \BibitemOpen
  \bibfield  {author} {\bibinfo {author} {\bibfnamefont {G.}~\bibnamefont
  {Brunello}}, \bibinfo {author} {\bibfnamefont {V.}~\bibnamefont {Chestnov}},
  \ and\ \bibinfo {author} {\bibfnamefont {P.}~\bibnamefont {Mastrolia}},\
  }\href@noop {} {\  (\bibinfo {year} {2024}{\natexlab{b}})},\ \Eprint
  {http://arxiv.org/abs/2408.16668} {arXiv:2408.16668 [hep-th]} \BibitemShut
  {NoStop}%
\bibitem [{\citenamefont {Henn}(2013)}]{hennAlg}%
  \BibitemOpen
  \bibfield  {author} {\bibinfo {author} {\bibfnamefont {J.~M.}\ \bibnamefont
  {Henn}},\ }\href {\doibase 10.1103/PhysRevLett.110.251601} {\bibfield
  {journal} {\bibinfo  {journal} {Phys. Rev. Lett.}\ }\textbf {\bibinfo
  {volume} {110}},\ \bibinfo {pages} {251601} (\bibinfo {year} {2013})},\
  \Eprint {http://arxiv.org/abs/1304.1806} {arXiv:1304.1806 [hep-th]}
  \BibitemShut {NoStop}%
\bibitem [{\citenamefont {Kotikov}\ \emph {et~al.}(2003)\citenamefont
  {Kotikov}, \citenamefont {Lipatov},\ and\ \citenamefont
  {Velizhanin}}]{lipKot}%
  \BibitemOpen
  \bibfield  {author} {\bibinfo {author} {\bibfnamefont {A.~V.}\ \bibnamefont
  {Kotikov}}, \bibinfo {author} {\bibfnamefont {L.~N.}\ \bibnamefont
  {Lipatov}}, \ and\ \bibinfo {author} {\bibfnamefont {V.~N.}\ \bibnamefont
  {Velizhanin}},\ }\href {\doibase 10.1016/S0370-2693(03)00184-9} {\bibfield
  {journal} {\bibinfo  {journal} {Phys. Lett. B}\ }\textbf {\bibinfo {volume}
  {557}},\ \bibinfo {pages} {114} (\bibinfo {year} {2003})},\ \Eprint
  {http://arxiv.org/abs/hep-ph/0301021} {arXiv:hep-ph/0301021} \BibitemShut
  {NoStop}%
\bibitem [{\citenamefont {Chestnov}\ \emph {et~al.}(2022)\citenamefont
  {Chestnov}, \citenamefont {Gasparotto}, \citenamefont {Mandal}, \citenamefont
  {Mastrolia}, \citenamefont {Matsubara-Heo}, \citenamefont {Munch},\ and\
  \citenamefont {Takayama}}]{Manosh}%
  \BibitemOpen
  \bibfield  {author} {\bibinfo {author} {\bibfnamefont {V.}~\bibnamefont
  {Chestnov}}, \bibinfo {author} {\bibfnamefont {F.}~\bibnamefont
  {Gasparotto}}, \bibinfo {author} {\bibfnamefont {M.~K.}\ \bibnamefont
  {Mandal}}, \bibinfo {author} {\bibfnamefont {P.}~\bibnamefont {Mastrolia}},
  \bibinfo {author} {\bibfnamefont {S.~J.}\ \bibnamefont {Matsubara-Heo}},
  \bibinfo {author} {\bibfnamefont {H.~J.}\ \bibnamefont {Munch}}, \ and\
  \bibinfo {author} {\bibfnamefont {N.}~\bibnamefont {Takayama}},\ }\href
  {\doibase 10.1007/JHEP09(2022)187} {\bibfield  {journal} {\bibinfo  {journal}
  {JHEP}\ }\textbf {\bibinfo {volume} {09}},\ \bibinfo {pages} {187} (\bibinfo
  {year} {2022})},\ \Eprint {http://arxiv.org/abs/2204.12983} {arXiv:2204.12983
  [hep-th]} \BibitemShut {NoStop}%
\bibitem [{\citenamefont {Chestnov}\ \emph {et~al.}(2023)\citenamefont
  {Chestnov}, \citenamefont {Matsubara-Heo}, \citenamefont {Munch},\ and\
  \citenamefont {Takayama}}]{Chestnov:2023kww}%
  \BibitemOpen
  \bibfield  {author} {\bibinfo {author} {\bibfnamefont {V.}~\bibnamefont
  {Chestnov}}, \bibinfo {author} {\bibfnamefont {S.~J.}\ \bibnamefont
  {Matsubara-Heo}}, \bibinfo {author} {\bibfnamefont {H.~J.}\ \bibnamefont
  {Munch}}, \ and\ \bibinfo {author} {\bibfnamefont {N.}~\bibnamefont
  {Takayama}},\ }\href {\doibase 10.1007/JHEP11(2023)202} {\bibfield  {journal}
  {\bibinfo  {journal} {JHEP}\ }\textbf {\bibinfo {volume} {11}},\ \bibinfo
  {pages} {202} (\bibinfo {year} {2023})},\ \Eprint
  {http://arxiv.org/abs/2305.01585} {arXiv:2305.01585 [hep-th]} \BibitemShut
  {NoStop}%
\bibitem [{\citenamefont {Brunello}\ \emph
  {et~al.}(2024{\natexlab{c}})\citenamefont {Brunello}, \citenamefont
  {Crisanti}, \citenamefont {Giroux}, \citenamefont {Mastrolia},\ and\
  \citenamefont {Smith}}]{Brunello:2023fef}%
  \BibitemOpen
  \bibfield  {author} {\bibinfo {author} {\bibfnamefont {G.}~\bibnamefont
  {Brunello}}, \bibinfo {author} {\bibfnamefont {G.}~\bibnamefont {Crisanti}},
  \bibinfo {author} {\bibfnamefont {M.}~\bibnamefont {Giroux}}, \bibinfo
  {author} {\bibfnamefont {P.}~\bibnamefont {Mastrolia}}, \ and\ \bibinfo
  {author} {\bibfnamefont {S.}~\bibnamefont {Smith}},\ }\href {\doibase
  10.1103/PhysRevD.109.094047} {\bibfield  {journal} {\bibinfo  {journal}
  {Phys. Rev. D}\ }\textbf {\bibinfo {volume} {109}},\ \bibinfo {pages}
  {094047} (\bibinfo {year} {2024}{\natexlab{c}})},\ \Eprint
  {http://arxiv.org/abs/2311.14432} {arXiv:2311.14432 [hep-th]} \BibitemShut
  {NoStop}%
\bibitem [{\citenamefont {Benincasa}\ \emph {et~al.}(2024)\citenamefont
  {Benincasa}, \citenamefont {Brunello}, \citenamefont {Mandal}, \citenamefont
  {Mastrolia},\ and\ \citenamefont {Vaz\~ao}}]{Benincasa:2024ptf}%
  \BibitemOpen
  \bibfield  {author} {\bibinfo {author} {\bibfnamefont {P.}~\bibnamefont
  {Benincasa}}, \bibinfo {author} {\bibfnamefont {G.}~\bibnamefont {Brunello}},
  \bibinfo {author} {\bibfnamefont {M.~K.}\ \bibnamefont {Mandal}}, \bibinfo
  {author} {\bibfnamefont {P.}~\bibnamefont {Mastrolia}}, \ and\ \bibinfo
  {author} {\bibfnamefont {F.}~\bibnamefont {Vaz\~ao}},\ }\href@noop {} {\
  (\bibinfo {year} {2024})},\ \Eprint {http://arxiv.org/abs/2408.16386}
  {arXiv:2408.16386 [hep-th]} \BibitemShut {NoStop}%
\end{thebibliography}%

\end{document}